\documentclass[aps,ams,amsmath,prx,longbibliography,twocolumn,superscriptaddress]{revtex4-2}
\usepackage{graphics}
\usepackage{graphicx}
\usepackage{epstopdf,epsfig}
\usepackage{amsmath}
\usepackage{amssymb}
\usepackage{amsfonts}
\usepackage{mathrsfs}
\usepackage{bm}
\usepackage{xcolor}
\usepackage{bbm}
\usepackage{hyperref}
\usepackage[normalem]{ulem}
\usepackage{comment}
\usepackage{soul}
\usepackage[varg]{txfonts}

\begin{document}

\title{Fluctuation-Induced Magnetoelectric Effect in Noncentrosymmetric Superconductors}

\author{Jaglul Hasan}
\affiliation{Department of Physics and Astronomy, Iowa State University, Ames, Iowa 50011, USA}
\affiliation{Ames National Laboratory, U.S. Department of Energy, Ames, Iowa 50011, USA}

\author{Daniel Shaffer}
\affiliation{Department of Physics, University of Wisconsin-Madison, Madison, Wisconsin 53706, USA}

\author{Maxim Dzero}
\affiliation{Department of Physics, Kent State University, Kent, Ohio 44242, USA}

\author{Alex Levchenko}
\affiliation{Department of Physics, University of Wisconsin-Madison, Madison, Wisconsin 53706, USA}

\date{June 24, 2026}

\begin{abstract}
We study the effect of superconducting fluctuations on the spin susceptibility and NMR relaxation rate in noncentrosymmetric two-dimensional materials above the superconducting transition temperature, considering arbitrary strength of impurity scattering. Employing a microscopic model with linear Rashba spin-orbit coupling, we show that superconducting fluctuations give rise to a direct contribution to the spin susceptibility through fluctuation-induced Cooper pairs. This fluctuation-driven magnetoelectric effect is possible even for purely $s$-wave singlet pairing, a mechanism that is forbidden in centrosymmetric systems. It competes with the reduction of the susceptibility below the Pauli value arising from the combined effects of the suppression of the density of states and quantum-interference localization processes. In contrast, superconducting fluctuations enhance the NMR relaxation rate above its normal-state Korringa value, with spin-orbit coupling providing an additional amplification of this effect.   
\end{abstract}

\maketitle

\section{Introduction}

In recent years, there has been a growing interest in the properties of noncentrosymmetric superconductors \cite{Mineev:99,Bauer:12}, stimulated by the discovery of robust superconductivity in transition-metal dichalcogenides (TMDs) such as $\text{NbSe}_2$ and $\text{MoS}_2$, as well as related moir\'e systems, including $\text{WTe}_2$, $\text{MoTe}_2$, and moir\'e-engineered twisted $\text{WSe}_2$, see Refs. \cite{Lu:15,Kang:15,Ugeda:16,Saito:16,Xi:16,Qi:16,Costanzo:16,Dvir:18,Barrera:18,Sohn:18,Cobden:18,Fatemi:18,Hamill:21,Xia:25}. Because these systems inherently lack an inversion center, they display a variety of fascinating nonreciprocal phenomena. Prominent examples include the superconducting diode effect (SDE) \cite{Nadeem:23,ma:25,Shaffer:25} and magnetochiral anisotropy (MCA) \cite{Tokura:18,Nagaosa:24}, both of which can be viewed as macroscopic manifestations of magnetoelectric phenomena unique to unconventional superconductors.

The observed giant enhancement of MCA \cite{Wakatsuki:17}, which manifests sharply at the onset of the superconducting transition and persists at lower temperatures, has revived significant interest in the role of superconducting fluctuations and their profound effects on both the transport and thermodynamic properties of materials \cite{Wakatsuki:18,Hoshino:18,Daido:24,Dong:25,JTM:26a,JTM:26b}. 

In this work, we focus specifically on the spin physics of these systems. In particular, we investigate the spin susceptibility and nuclear magnetic resonance (NMR) spin-lattice relaxation rate within the regime of strong superconducting fluctuations in materials lacking an inversion center. Within linear response theory, both of these physical quantities can be  extracted from the electronic spin-spin correlation function via the Kubo formula. Technically, the NMR relaxation rate, denoted as $1/T_1$, quantifies how quickly a nuclear spin system returns to thermal equilibrium with its surrounding lattice environment following a magnetic perturbation. According to the Bloch-Wangsness-Redfield theory \cite{Bloch:53,Redfield:57}, this relaxation rate is driven by time-dependent, fluctuating local magnetic fields experienced by the nuclei, which typically originate from the hyperfine coupling to neighboring electronic spins. Using the hyperfine interaction Hamiltonian, $1/T_1$ can be rewritten directly in terms of the dynamic electronic spin-spin correlation function. 
For a normal, non-interacting Fermi liquid, this correlation function is dominated by low-energy electron-hole excitations near the Fermi surface, leading to a characteristic linear temperature dependence known as the Korringa relation \cite{Korringa:50}. Below the superconducting transition temperature $T_c$, however, the opening of a pairing gap $\Delta$ fundamentally alters the electronic spin-spin correlation function. In conventional fully gapped $s$-wave superconductors, this causes $1/T_1$ to drop exponentially at low temperatures as $\sim e^{-\Delta/T}$, often immediately following a distinct, fluctuation-enhanced Hebel-Slichter coherence peak just below $T_c$ \cite{HebelSlichter:57}. Conversely, in unconventional superconductors featuring a nodal gap structure, where the superconducting gap vanishes at specific points or lines on the Fermi surface, the spin-lattice relaxation rate exhibits a power-law temperature dependence, $1/T_1 \propto T^n$, at low temperatures, $T \ll T_c$, rather than exponential suppression. The specific power-law exponent $n$ depends directly on the geometry and dimensionality of these nodes.

The primary source of anomalous responses near $T_c$ is related to the formation of fluctuation-driven Cooper pairs \cite{LarkinVarlamov:05}. It is well established that there are three distinct microscopic mechanisms through which these fluctuating Cooper pairs affect physical quantities. These are the density of states (DOS) effect \cite{Abrahams:70}, the Aslamazov-Larkin (AL) process associated with paraconductivity \cite{AL:68}, and the Maki-Thompson (MT) process stemming from quantum interference effects in the Cooper channel \cite{Maki:68,Thompson:70}. The original works on this topic addressed the problem of NMR in both clean and disordered superconductors near $T_c$ \cite{Maniv:76,Kuboki:89,Heym:92}.  The corresponding theoretical calculations for the spin susceptibility within these regimes were later systematically carried out in Ref. \cite{Randeria:94}. 
Subsequently, these results were generalized to map out the entire phase diagram of fluctuations across a wide range of magnetic fields and temperatures, specifically extending into the quantum regime near the upper critical field $H_{c2}$ \cite{Glatz:15}. Alongside these advancements, we also highlight related studies focusing on the non-analytical corrections to spin susceptibility within Fermi liquid theory \cite{BKM:93,Belitz:97,Chitov:01,Chubukov:04,Betouras:05}. In the following, we briefly recapitulate the key results from these important studies in order to properly contextualize our work within the broader field.

In the vicinity of the critical temperature $T_c$, where $\epsilon=\ln(T/T_c)\approx (T-T_c)/T_c\ll 1$, the relative importance of the AL, MT, and DOS contributions to the spin susceptibility and NMR relaxation rate differs from their interplay in other quantities, such as the electrical conductivity. This is primarily due to the fact that the usually dominant AL contribution is absent in the spin-spin correlation function for $s$-wave pairing. Indeed, the spin operator at the vertex of the Kubo response function flips the spin, and therefore the electron Green's functions adjacent to the vertex must carry opposite spins. The AL diagram contains a triangular vertex, and there is no consistent way to assign a spin to the third electron Green's function so as to form a pair propagator in the singlet channel. It turns out that the anomalous MT contribution is also absent \cite{Randeria:94}.

As a result, superconducting fluctuations suppress the spin susceptibility through the combined effects of the depletion of the density of states near the Fermi level and the regular part of the MT process. In the disordered limit, quantum-interference diagrams (namely, Cooperon impurity ladders, which play a crucial role in weak localization) determine the sign of the correction to the spin susceptibility.

In contrast, the anomalous MT contribution leads to an enhancement of $1/(T_1T)$ above its normal-state Korringa value. In particular, superconducting fluctuations above $T_c$ produce an effect opposite in sign to that observed deep in the superconducting state ($T\ll T_c$), where $1/T_1$ decreases exponentially with temperature. Strong dephasing further suppresses the anomalous MT contribution, in which case the NMR relaxation rate is governed by the less singular DOS and regular MT terms.

In the clean limit, $T\tau\gg 1$, where $\tau$ is the elastic impurity-scattering time, there is a peculiar crossover in the temperature dependence of the NMR relaxation rate, depending on whether $T_c\tau$ is larger or smaller than $1/\sqrt{\epsilon}$.

Finally, we note that even for repulsive interactions in a Fermi-liquid regime, rescattering of quasiparticle pairs in the Cooper channel leads to a strong renormalization of the nonanalytic corrections to the spin susceptibility \cite{Schwiete:06,Shekhter:06}.

We now return to this problem in the context of noncentrosymmetric superconductors. These systems exhibit pronounced spin-orbit effects and, in the simplest case, can be described by Rashba spin-orbit coupling \cite{Rashba:59,Bychkov:84}. This interaction generates spin-split Fermi surfaces with a characteristic spin-momentum-locked helical spin texture. One of the most profound manifestations of Rashba spin-orbit coupling in conductors is the magnetoelectric effects (MEE) predicted by Levitov-Nazarov-Eliashberg \cite{LNE:85a,LNE:85b} and Edelstein \cite{Edelstein:95,Edelstein:05}, whereby a spin polarization of the charge carriers is induced by an electric field in the normal state or supercurrent in the superconductor. Another important consequence is the emergence of a triplet component of the anomalous (off-diagonal) correlation function in addition to the singlet component, even when the superconducting order parameter itself is purely singlet. Spin susceptibility and NMR in superconductors without inversion symmetry below $T_c$ were calculated in Refs. \cite{Frigeri:04,Hayashi:06}.  
We show that, above $T_c$, fluctuations of the supercurrent are accompanied by fluctuations of the spin polarization. The resulting finite mean-square spin polarization gives rise to a correction to the spin susceptibility through the AL process, which, as discussed above, is forbidden in centrosymmetric systems. We further demonstrate that the same mechanism leads to additional fluctuation corrections to the NMR relaxation rate. We study this fluctuation-driven magnetoelectric effect as a function of the spin-orbit coupling strength and show that it survives in disordered systems, although it gets  progressively suppressed by impurity scattering.

%@@@@@@@@@@@@@@@@@@@@@@@@@@@@@@@@@@@@@@@@@@@@@@@@@@@@@@@@@@@@@@@@@@@@@@@@@@@@@@@@@@@@@
\begin{figure}
\includegraphics[width=0.45\textwidth]{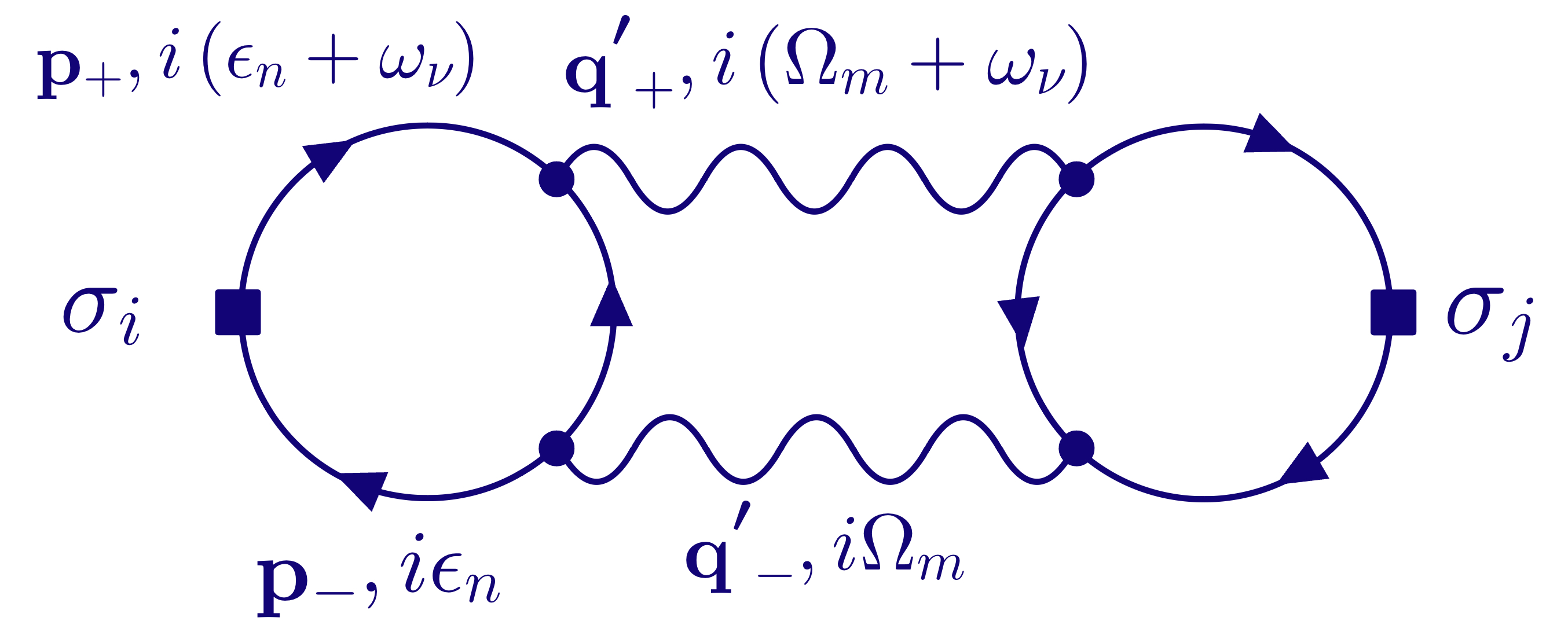} 
\centering \caption{Aslamazov-Larkin (AL) diagram for the spin-spin correlation function. The spin vertex in each block is denoted by a box with spin $\sigma_{i,j}$. The wavy lines describe the pair-propagator, and straight lines correspond to the single-particle electron Green's function. Here, $\mathbf{p_\pm}=\mathbf{p} \pm \frac{\mathbf{q}}{2}$ and $\mathbf{q^{\prime}_\pm}=\mathbf{q^{\prime}} \pm \frac{\mathbf{q}}{2}$. The external frequency is denoted by $\omega_\nu$ and fermionic Matsubara energies are denoted by $\epsilon_n$.}\label{fig:AL}
\end{figure}
%@@@@@@@@@@@@@@@@@@@@@@@@@@@@@@@@@@@@@@@@@@@@@@@@@@@@@@@@@@@@@@@@@@@@@@@@@@@@@@@@@@@@@

%$$$$$$$$$$$$$$$$$$$$$$$$$$$$$$$$$$$$$$$$$$$$$$$$$$$$$$$$$$$$$$$$$$$$$
%$$$$$$$$$$$$$$$$$$$$$$$$$$$$$$$$$$$$$$$$$$$$$$$$$$$$$$$$$$$$$$$$$$$$$
%$$$$$$$$$$$$$$$$$$$$$$$$$$$$$$$$$$$$$$$$$$$$$$$$$$$$$$$$$$$$$$$$$$$$$
\section{Model and Technique}

To describe the systems under consideration and perform explicit calculations, we employ the Edelstein model \cite{Edelstein:05,Edelstein:21}. The model incorporates several essential ingredients: a two-dimensional (2D) electron system with an isotropic parabolic dispersion, linear Rashba spin-orbit coupling (SOC) term, a static scalar disorder potential, and conventional s-wave BCS pairing. The corresponding Hamiltonian is given by
\begin{align}\label{eq:H}
H=\int d^2\mathbf{r}&\left[\psi^\dagger_a(\mathbf{r})\left(-\frac{\nabla^2}{2m}-\mu+V(\mathbf{r})\right)\psi_a(\mathbf{r})\right. \nonumber \\ 
 &-i\alpha_R\psi^\dagger_a(\mathbf{r})([\nabla\times\mathbf{c}]\cdot\bm{\sigma}_{ab}) \psi_b(\mathbf{r}) \nonumber \\ 
 &+\left. \frac{\lambda_s}{4}\left[\psi_a^{\dagger}(\mathbf{r}) g_{ab} \psi_b^{\dagger}(\mathbf{r})\right]\left[\psi_c(\mathbf{r}) g_{cd}^{t} \psi_d(\mathbf{r})\right]\right], 
\end{align}
where $\psi^\dagger(\mathbf{r})$ and $\psi(\mathbf{r})$ are the usual fermion creation and annihilation field operators. The single-particle part of the above Hamiltonian of the clean systems has the form upon the Fourier transform $\frac{\mathbf{p}^2}{2 m}-\mu+\alpha_R([\mathbf{p} \times \mathbf{c}] \cdot \bm{\sigma})$. Here $\mathbf{p}$ is the momentum, $m$ is the electron effective mass, $\mathbf{c}$ is a unit vector perpendicular to the 2D system, while $\bm{\sigma}$ is the Pauli spin matrix, and $\alpha_{R}$ is the strength of the Rashba spin-orbit interaction with dimensions of velocity (we work in the natural units by setting $\hbar=1$ and $k_B=1$).
The spin-orbit coupling lifts the spin degeneracy of the conduction electrons, giving rise to two energy bands of positive and negative helicities \(\lambda=\pm1\)
with energies $E_{\lambda}(p)=\frac{p^2}{2 m} -\mu+\lambda \alpha_{R} p$. 

We include disorder as a short-ranged potential of impurities placed at randomly distributed points $\mathbf{R}_i$ with concentration $n_{\text{imp}}$: 
$V(\mathbf{r})=\sum_i U \delta\left(\mathbf{r}-\mathbf{R}_i\right)$, where $U$ is the potential of individual impurity center. For these scatterers, 
the elastic scattering time $\tau$ can be computed from the leading Born approximation, giving $\tau^{-1}=2 \pi \nu n_{\text{imp}}|U|^2$, where $\nu=m / 2 \pi$ is the density of states on the Fermi level.

We assume in Eq. \eqref{eq:H} s-wave singlet pairing interaction to be the only interaction between the electrons:
where $\hat{g}=i\hat{\sigma}^y$, $\lambda_s$ is the pairing constant, and the superscript-$\mathrm{t}$ in $g^t_{cd}$ denotes matrix transposition.

Working at temperatures close to the transition temperature $T_c$, there are four independent dimensionless parameters in this model:
\begin{equation}\label{eq:parameters}
\epsilon=\ln \frac{T}{T_c}, \quad \delta=\frac{\alpha_R }{v_F}, \quad \varkappa=\alpha_R p_F\tau, \quad \varrho=\pi T\tau,
\end{equation}
where $p_F=\sqrt{2m\mu}$ is the Fermi momentum at chemical potential $\mu$, and $v_F=p_F/m$ is the Fermi velocity (in the absence of SOC). The parameter $\epsilon =\ln \frac{T}{T_c} \approx \frac{T-T_c}{T_c}$ is the smallest close to the transition temperature $T_c$. The parameter $\delta$ is also assumed to be small, but the parameters $\varkappa$ and $\varrho$ are arbitrary. 

The pairing propagator $g_{ab}L\left(\mathbf{q}, \Omega_m\right)g^{\mathrm{t}}_{cd}$ (wavy line in Fig. \ref{fig:AL}) describes pairing fluctuations above $T_c$, where $\mathbf{q}$ is the collective momentum and $\Omega_m$ is the Matsubara frequency of the virtual Cooper pair. The spinor indices $ab$ and $cd$ refer to the spin states of the carriers joining to form the pair and releasing after its subsequent decay, respectively. Since we limit our attention to temperatures close to the transition temperature when $\epsilon\ll 1$, it suffices to focus on long-wavelength, linear in $\Omega_m=2\pi m T$ fluctuations. 
Thus we have for $L\left(\mathbf{q}, \Omega_m\right)$:
\begin{equation}\label{eq:flu_prop}
L^{-1}\left(\mathbf{q}, \Omega_m\right)=-\nu_d\left[\epsilon+\frac{\pi |\Omega_m|}{8T}+\eta_d q^2\right],
\end{equation}
up to corrections of the order of $\sim\left(\alpha_R /v_F\right)^2\ll1$. Here, the subscript $d$ denotes the dimensionality of the system, for example, in 2D systems, the density of states per one spin, $\nu_2=\nu=m/2\pi$ and
\begin{equation}\label{eq:eta2}
\begin{aligned}
\eta_{2} & =-\frac{v_F^2 \tau^2}{2}\biggl[\psi\left(\frac{1}{2}+\frac{1}{4 \pi T\tau}\right)-\psi\left(\frac{1}{2}\right)-\frac{1}{4 \pi T \tau} \psi^{(1)}\left(\frac{1}{2}\right)\biggr] \\
& \approx \begin{cases}\pi D /\left(8 T_c\right) & \text { for } T_c \tau \ll 1, \\
7 \zeta(3) v_F^2 /\left(32 \pi^2 T_c^2\right) & \text { for } T_c \tau \gg 1,\end{cases}
\end{aligned}
\end{equation}
turns out to be proportional to the square of effective coherence length, $\xi_2=\eta_2^{\frac{1}{2}}$. Here, $D=v_F^2\tau/2$ is the 2D diffusion constant, $\psi(x)$, $\zeta(x)$ are the digamma and Riemann zeta functions, respectively with $\psi^{(N)}(x)$ being the $N$-th order derivative of the digamma function. 

%@@@@@@@@@@@@@@@@@@@@@@@@@@@@@@@@@@@@@@@@@@@@@@@@@@@@@@@@@@@@@@@@@@@@@@@@@@@@@@@@@@@@@
\begin{figure}
\includegraphics[width=0.45\textwidth]{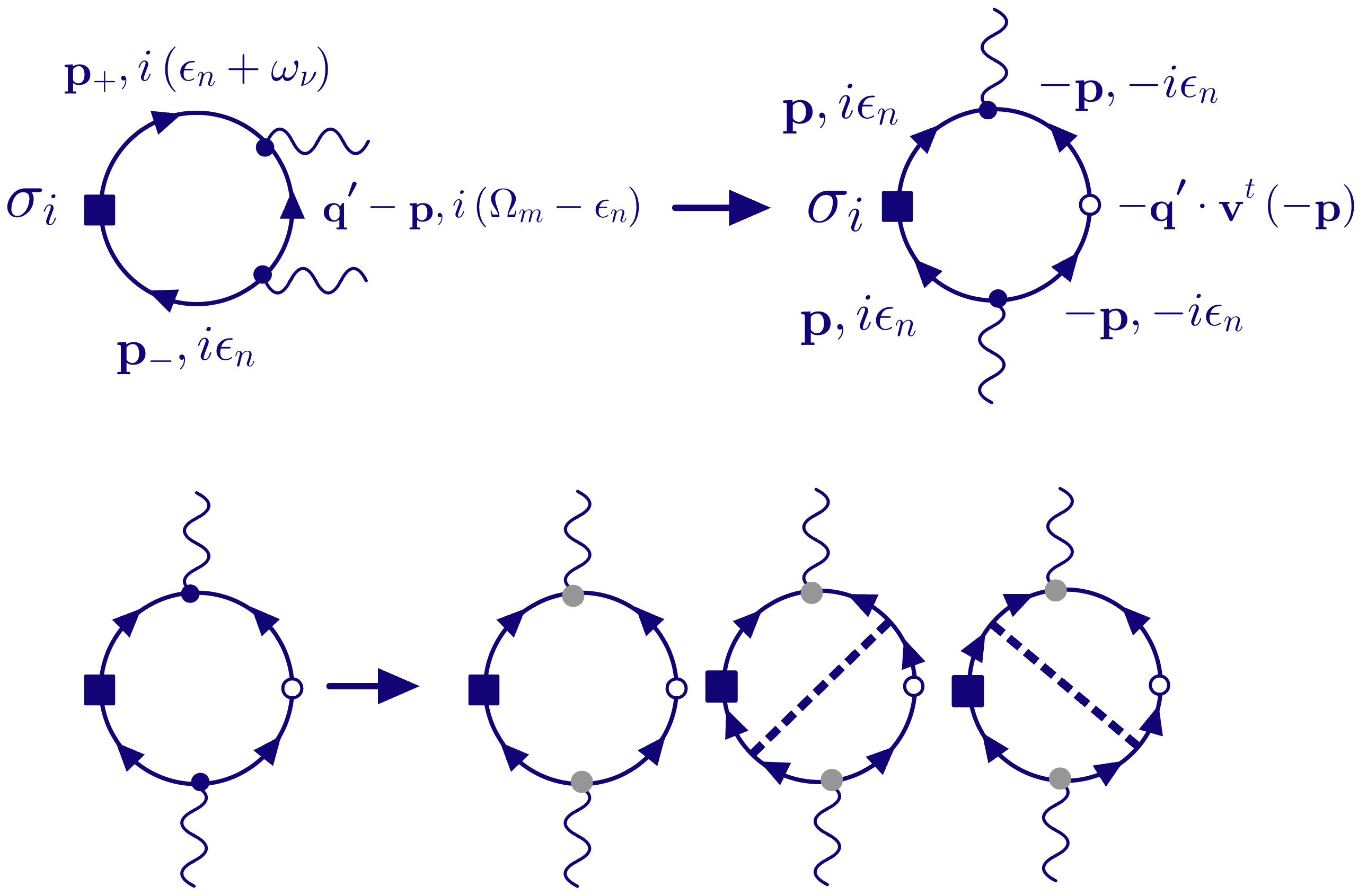} 
\centering
\caption{In the top panel, the diagram on the left represents the block $B_i(\mathbf{q}',\Omega_m,\omega_\nu)$ appearing on the left side of the AL diagram in Fig.~\ref{fig:AL}. After expanding to linear order in $\mathbf{q}'$, it is transformed into the diagram shown on the right. In the bottom panel, the leftmost diagram represents the skeletal diagram in the clean limit. Averaging over random impurity realizations generates the three diagrams shown on the right.} \label{fig:Block}
\end{figure}
%@@@@@@@@@@@@@@@@@@@@@@@@@@@@@@@@@@@@@@@@@@@@@@@@@@@@@@@@@@@@@@@@@@@@@@@@@@@@@@@@@@@@@

In the following $B_{i,j}\left(\mathbf{q}^\prime, \Omega_m, \omega_{\nu}\right)$ denotes the right and left circular blocks of the AL diagram in Fig. \ref{fig:AL} containing the spin vertices, shown diagrammatically in Fig. \ref{fig:Block}, integrated over the electron loop momentum and summed over the fermionic frequency. In the vicinity of $T_c$, we can neglect the $\Omega_m$ and $\omega_{\nu}$-dependencies of $B_{i,j}\left(\mathbf{q}^\prime, \Omega_m, \omega_{\nu}\right)$ due to the pole structure of the fluctuation propagator in Eq. (\ref{eq:flu_prop}). These diagrams are evaluated by standard diagram technique in Appendix \ref{AppA} and we only give the final expression here: 
\begin{equation}\label{eq:block}
B_{i,j}(\mathbf{q}^\prime) =\sum_{\omega_n>0} \frac{4\pi \nu \alpha_R  T\tau \varkappa^2\left[\mathbf{c} \times \mathbf{q}^\prime\right]_{i,j}}{\omega^2_n[\tau \omega_n\left(2 \tau \omega_n+1\right)^2+\varkappa^2\left(4 \tau \omega_n+1\right)]}.
\end{equation}
Owing to spin-orbit coupling, this vertex becomes proportional to the collective Cooper pair momentum similar as it happens in the AL diagram for the electrical conductivity.  
One might naively expect this correction to the spin susceptibility to exhibit the same dependence on $\epsilon$ as the paraconductivity in the conventional case. This is, however, not the case. Since the spin susceptibility is a thermodynamic quantity, its leading singular behavior in $\epsilon$ can be obtained by evaluating the response function at zero Matsubara frequency. In contrast, the conductivity must be extracted from the properly analytically continued response function. Taking the dc limit then requires differentiation of the pair propagator with respect to frequency, resulting in a higher-order pole and, consequently, a stronger singularity.     

%@@@@@@@@@@@@@@@@@@@@@@@@@@@@@@@@@@@@@@@@@@@@@@@@@@@@@@@@@@@@@@@@@@@@@@@@@@@@@@@@@@@@@
\begin{figure}
\includegraphics[width=0.45\textwidth]{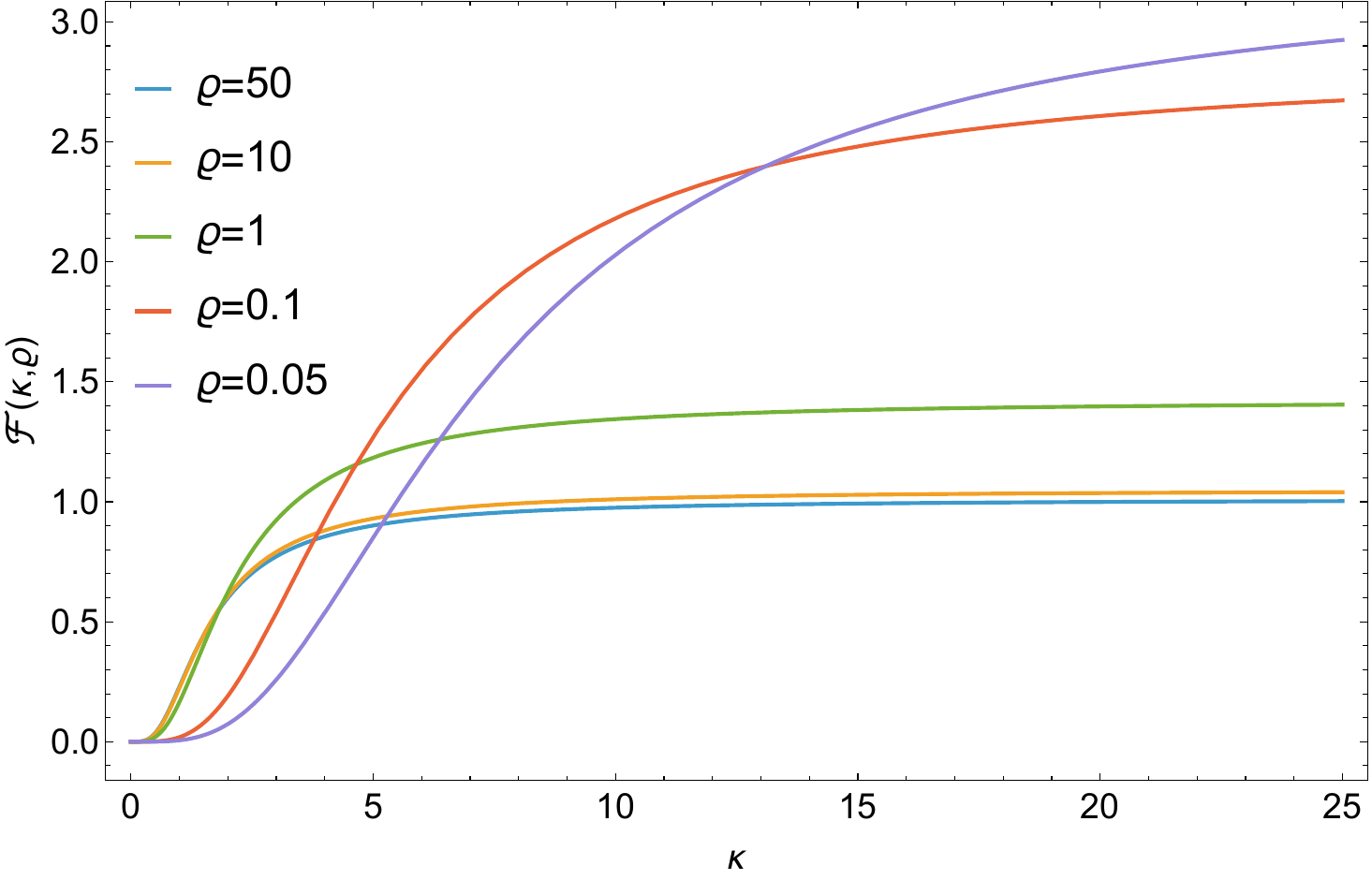}
\includegraphics[width=0.45\textwidth]{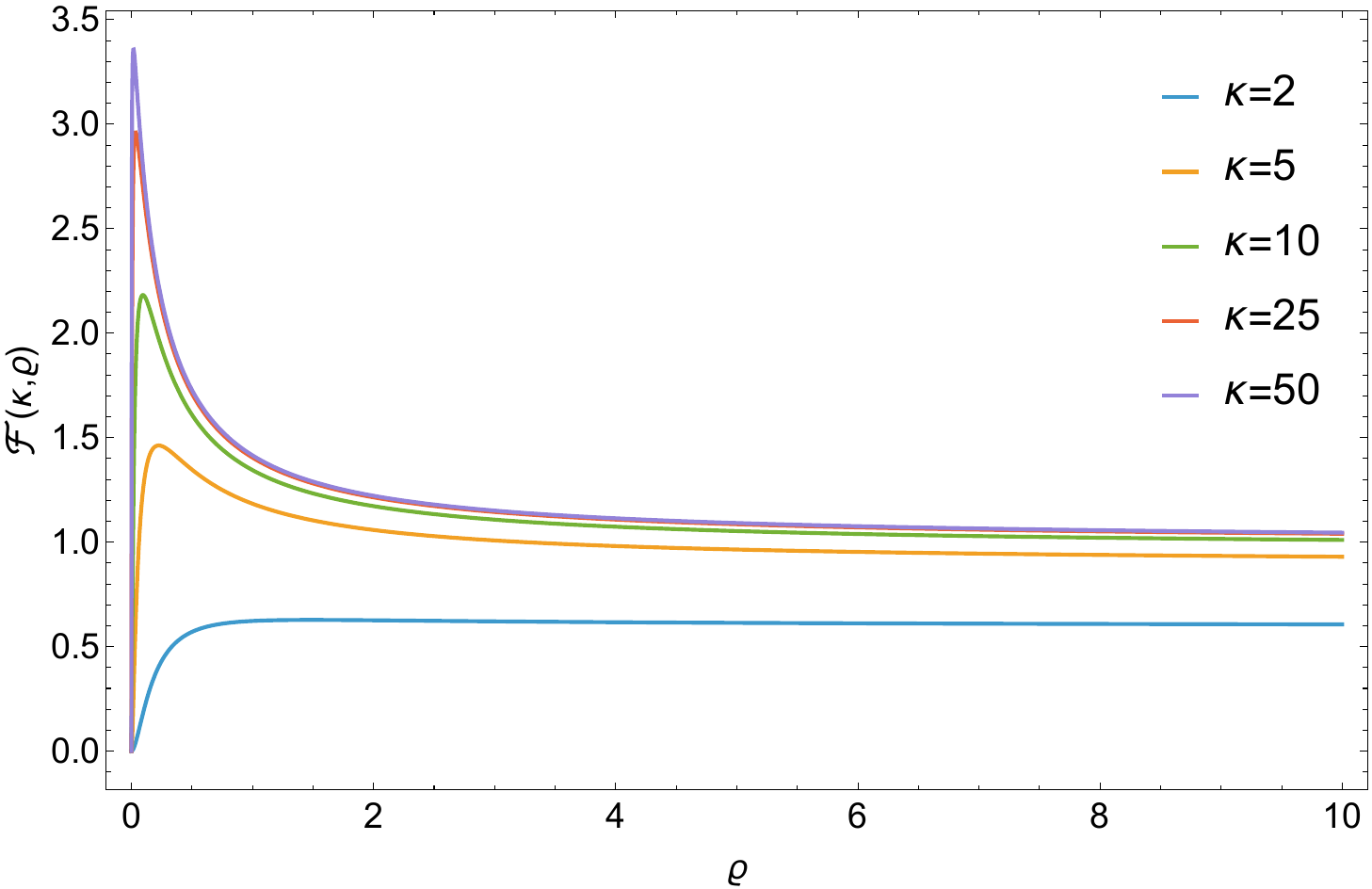}
\centering \caption{The upper panel shows $\mathcal{F}$ as a function of the spin-orbit coupling strength, $\kappa=\alpha_R p_F/\pi T$, for different values of the disorder parameter $\varrho=\pi T\tau$. The ballistic regime corresponds to $\varrho>1$, while the diffusive regime corresponds to $\varrho<1$. The lower panel shows the same function plotted as a function of $\varrho$ for several values of $\kappa$.}\label{fig:F}
\end{figure}
%@@@@@@@@@@@@@@@@@@@@@@@@@@@@@@@@@@@@@@@@@@@@@@@@@@@@@@@@@@@@@@@@@@@@@@@@@@@@@@@@@@@@@

%$$$$$$$$$$$$$$$$$$$$$$$$$$$$$$$$$$$$$$$$$$$$$$$$$$$$$$$$$$$$$$$$$$$$$
%$$$$$$$$$$$$$$$$$$$$$$$$$$$$$$$$$$$$$$$$$$$$$$$$$$$$$$$$$$$$$$$$$$$$$
%$$$$$$$$$$$$$$$$$$$$$$$$$$$$$$$$$$$$$$$$$$$$$$$$$$$$$$$$$$$$$$$$$$$$$
\section{Spin susceptibility}

The computation for the fluctuation-induced correction to spin susceptibility can be made simpler by setting the external frequency and momentum to zero at the beginning, which gives 
\begin{equation}\label{eq:spin_sus}
\delta\chi_{ij}=T \int_{\mathbf{q}} L^2(\mathbf{q}) B_i\left(\mathbf{q}\right) B_j\left(\mathbf{q}\right),
\end{equation}
where $\int_{\mathbf{q}}=\int \frac{d^2 \mathbf{q}}{(2 \pi)^2}=\int \frac{d \hat{q}}{2 \pi}\int \frac{q dq}{2 \pi}$. After substituting Eq. \eqref{eq:flu_prop} together with Eq. \eqref{eq:block} in Eq. \eqref{eq:spin_sus}, we are left to do the following angular and momentum integrals:
\begin{subequations}
\begin{equation}\label{eq:q-integral-phi}
\int \frac{d \hat{q}}{2 \pi}\left[\mathbf{c} \times \mathbf{q}\right]_{i}\left[\mathbf{c} \times \mathbf{q}\right]_{j}=\frac{1}{2}\left(\delta_{ij}-c_i c_j\right)q^2,
\end{equation}
\begin{equation}
\int_0^{1/\xi} \frac{q^{3} dq}{2 \pi} \frac{1}{\left(\epsilon+\eta_2 q^2\right)^2}\approx\frac{1}{4\pi\eta_2^2}\ln \left(\frac{1}{\epsilon}\right),
\end{equation}
\end{subequations}
where we have introduced the standard cut-off $q_{\max} \sim \xi^{-1}=\eta_{2}^{-1 / 2}$ in order to regularize the ultraviolet divergence in the $q$-integration. Finally, we can get the final expression for the fluctuation-induced spin-susceptibility via AL process 
\begin{align}\label{eq:spin-sus-general}
& \delta\chi_{ij} =\left(\delta_{ij}-c_i c_j\right)\delta\chi,\nonumber\\
&\frac{\delta\chi}{\chi_{P}}=\frac{T}{\varepsilon_F} \ln \left(\frac{1}{\epsilon}\right)\left(\frac{\alpha_R}{v_F}\right)^2\mathcal{F}\left(\alpha_R p_F \tau,\pi T \tau\right),  
\end{align}
where $\chi_P=2\nu$ is the Pauli susceptibility. The dimensionless two-parameter function $\mathcal{F}(\varkappa,\varrho)$ can be calculated analytically in the clean limit, when $\tau\gg\max\left(1/T_c,1/\alpha_R p_F\right)$. We find
\begin{align}\label{eq:spin-sus-clean}
&\mathcal{F}(\varkappa,\varrho)= \left[1-\frac{4}{7\zeta(3) \kappa^2} \operatorname{Re}\left(\psi\left(\frac{1}{2}+\frac{i \kappa}{2}\right)-\psi\left(\frac{1}{2}\right)\right)\right]^2\nonumber \\
&\approx \begin{cases} 1-\frac{0.95}{\kappa^2} \left[\ln \frac{\kappa}{2}-\psi\left(\frac{1}{2}\right)\right] & \text { for } \kappa \gg 1 \text {  (strong SOC)}, \\
0.91 \kappa^4 & \text { for } \kappa \ll 1 \text {  (weak SOC)},\end{cases}
\end{align}
where we introduced the ratio $\kappa=\frac{\varkappa}{\varrho}=\frac{\alpha_R p_F}{\pi T}$, which characterizes the SOC strength. 

The dirty limit of $\mathcal{F}$ can also be analytically studied in two limiting cases of strong and weak SOC. In the strong SOC regime $(\kappa \gg 1)$, when $\varkappa \gg 1 \gg \varrho$ or, equivalently, $\alpha_R p_F \gg \tau^{-1} \gg T_c$, we find,
\begin{equation}\label{eq:spin-sus-dirty-strongSOC}
\mathcal{F}(\varkappa,\varrho)=\left[2+\frac{16\varrho}{\pi^2}\left(\ln(16\varrho)+\psi\left(\frac{1}{2}\right)\right)\right]^2 \approx 4.
\end{equation}
Notice that the fluctuation-driven spin susceptibility correction of the dirty limit in the strong SOC regime becomes $4$ times as big as that in the clean limit because of the disorder-induced mixing of singlet and triplet pairing correlations. In the weak SOC regime $(\kappa \ll 1)$, when $\varkappa \ll \varrho \ll 1$ or, equivalently, $\alpha_R p_F \ll T_c \ll \tau^{-1}$, we find:
\begin{equation}\label{eq:spin-sus-dirty-weakSOC}
\mathcal{F}(\varkappa,\varrho)\approx \frac{196\zeta^2(3)}{\pi^4} \kappa^4 \varrho^2.
\end{equation}
The full form of the $\mathcal{F}$-function is plotted in Fig. \ref{fig:F} for various parameter values across the ballistic-to-diffusive crossover at different strengths of the spin-orbit splitting.  

As a result, we find that the correction $\delta\chi$ arising from the fluctuation-induced MEE has the opposite sign to the DOS and regular MT contributions to the spin susceptibility \cite{Randeria:94}. Importantly, because this effect requires broken inversion symmetry, it carries an additional prefactor $(\alpha_R/v_F)^2$, where the two powers originate from the two triangular blocks entering the AL process. 
The suppression factor for this effect in the diffusive limit is sufficiently strong, introducing an additional small parameter proportional to $(T_c\tau)^2 \ll 1$.

%$$$$$$$$$$$$$$$$$$$$$$$$$$$$$$$$$$$$$$$$$$$$$$$$$$$$$$$$$$$$$$$$$$$$$
%$$$$$$$$$$$$$$$$$$$$$$$$$$$$$$$$$$$$$$$$$$$$$$$$$$$$$$$$$$$$$$$$$$$$$
%$$$$$$$$$$$$$$$$$$$$$$$$$$$$$$$$$$$$$$$$$$$$$$$$$$$$$$$$$$$$$$$$$$$$$
\section{NMR relaxation rate}

The calculation of the NMR relaxation rate is more involved as it requires the knowledge of the imaginary part of the dynamic spin susceptibility, $\operatorname{Im} \chi_{ij}^{(R)}(\mathbf{q}, \omega)$. The dimensionless relaxation rate of interest, $K=1/(T_1T)$, normalized to temperature, is defined by the relation  
\begin{equation}\label{eq:K}
K=\lim _{\omega \rightarrow 0} \frac{A}{\omega} \int_{\mathbf{q}} \operatorname{Im} \chi_{+-}^{(R)}(\mathbf{q}, \omega),
\end{equation}
where the constant $A>0$ comes from the microscopic hyperfine interaction. For noninteracting electrons at low temperatures, $T\ll \varepsilon_F$, $K=4\pi A\nu^2$, which is the Korringa constant. 
Because of the subtleties of the analytic continuation, this calculation of $K$ requires somewhat more care than the evaluation of $\delta \chi$.

%@@@@@@@@@@@@@@@@@@@@@@@@@@@@@@@@@@@@@@@@@@@@@@@@@@@@@@@@@@@@@@@@@@@@@@@@@@@@@@@@@@@@@
\begin{figure} 
\includegraphics[width=0.45\textwidth]{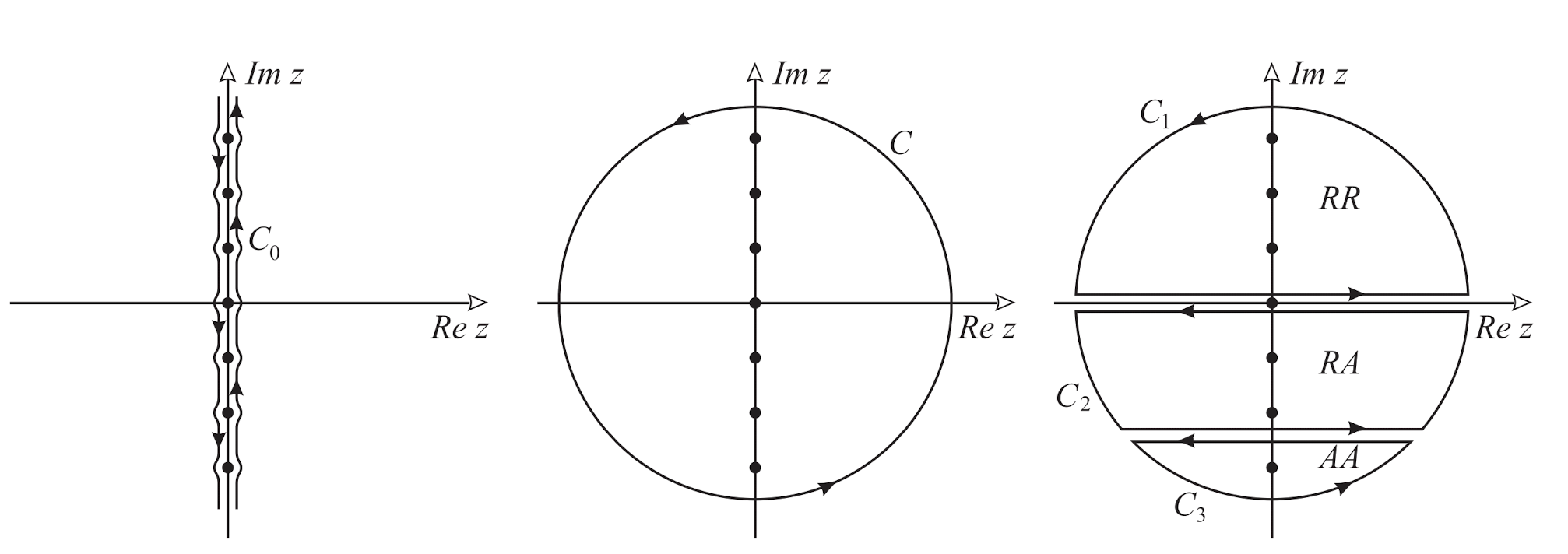} 
\centering
\caption{The integration contour in the complex frequency plane.}  \label{fig:contours}
\end{figure}
%@@@@@@@@@@@@@@@@@@@@@@@@@@@@@@@@@@@@@@@@@@@@@@@@@@@@@@@@@@@@@@@@@@@@@@@@@@@@@@@@@@@@@

The AL contribution to dynamic susceptibility $\chi_{ij}^{(R)}(\mathbf{q}, \omega) =\chi_{ij}\left(\mathbf{q}, i \omega_\nu \rightarrow \omega+i 0^{+}\right)$ can be calculated from the diagram in Fig. \ref{fig:AL} by performing the following $\Omega_m$ summation:
\begin{equation}
S=T \sum_m L\left(\mathbf{q}_{-}^{\prime}, \Omega_m\right) L\left(\mathbf{q}_{+}^{\prime}, \Omega_m+\omega_\nu\right),
\end{equation}
followed by an analytical continuation from the discrete set of frequencies to continuous variable in the complex plane. 
In order to complete these steps we convert the summation over bosonic Matsubara frequencies $\Omega_m$ into a contour integral:
\begin{equation}
T \sum_{\Omega_m} f\left(\Omega_m\right)=\frac{1}{4 \pi i} \oint_{C_0} d z \operatorname{coth} \frac{z}{2 T} f(-i z),
\end{equation}
where all bosonic Matsubara frequencies across which the summation is performed are enclosed by the contour $C_0$, and $z=i \Omega_m$ is a variable in the plane of complex frequency. The result is 
\begin{equation}
S=\frac{1}{4 \pi i} \oint_C d z \operatorname{coth} \frac{z}{2 T} L(\mathbf{q}_{-}^{\prime},-i z) L\left(\mathbf{q}_{+}^{\prime},-i z+\omega_\nu\right)
\end{equation}
where we choose the integration contour $C$ to represent a continuously deformed contour $C_0$ for ease of future integration.
With the next steps we follow the textbook procedure which mirrors that in the calculation of the conductivity diagram \cite{LarkinVarlamov:05}.  
The integrand above exhibits branches of nonanalyticity along the lines $\operatorname{Im} z=0$ and $\operatorname{Im} z=-i \omega_\nu$. Although the fluctuation propagator $L\left(\mathbf{q}, \Omega_k\right)$ is strictly defined only at discrete bosonic Matsubara frequencies, we must now evaluate it as a function of the continuous complex variable $z$. The analytic properties of Green's functions in the complex $z$-plane allow us to introduce two distinct analytic functions related to $L\left(\mathbf{q}, \Omega_k\right)$. The first is the retarded propagator, $L^R(\mathbf{q},-i z)$, which is analytic in the upper half-plane $(\operatorname{Im} z>0)$; the second is the advanced propagator, $L^A(\mathbf{q},-i z)$, which is devoid of singularities in the lower half-plane $(\operatorname{Im} z<0)$. Consequently, the $z$-plane can be sectioned via branch cuts along the lines $\operatorname{Im} z=0$ and $\operatorname{Im} z=-i \omega_\nu$. This allows us to choose the integration contour $C$ as the sum of three closed loops, $C_1+C_2+C_3$ (see Fig. \ref{fig:contours}), each bounding a domain where the integrand is strictly analytic. Integrating along each of the cuts, shifting the integration variable in one of the integrals, $z=z'-i\omega_\nu$, and considering that $i\omega_\nu$ is the period of $\coth\frac{z}{2T}$, we finally get an expression analytic in $i\omega_\nu\to\omega$  
\begin{align}
S=&\frac{1}{2 \pi} \int_{-\infty}^{+\infty} d z \operatorname{coth}\left(\frac{z}{2 T}\right) \nonumber \\ 
&\times\left[L^R\left(\mathbf{q}_{+}^{\prime},-i z-i\omega\right) \operatorname{Im} L^R\left(\mathbf{q}_{-}^{\prime},-i z\right)\right. \nonumber \\
 &\left.+ L^{R*}\left(\mathbf{q}_{-}^{\prime},-i z+i\omega\right) \operatorname{Im} L^R\left(\mathbf{q}_{+}^{\prime},-i z\right)\right].
\end{align}
We now use the explicit form of properly analytically continued pair propagator
\begin{equation}
L^R\left(\mathbf{q}, -i z\right)=-\frac{1}{\nu} \frac{1}{\epsilon-\frac{i\pi z}{8 T}+\eta_2 q^2}.
\end{equation}
To get the imaginary part contributing to Eq. \eqref{eq:K}, we only need the first order in $\omega$ term of the sum, which is 
\begin{equation}
S=\frac{\nu^{-2} i \omega}{8 T} \int_{-\infty}^{+\infty} d z \frac{\left(\frac{\pi z}{8 T}\right)\operatorname{coth}\left(\frac{z}{2 T}\right)\left[Q^2_+Q^2_--\left(\frac{\pi z}{8 T}\right)^4\right]}{\left[Q^2_-+\left(\frac{\pi z}{8 T}\right)^2\right]^2\left[Q^2_++\left(\frac{\pi z}{8 T}\right)^2\right]^2}
\end{equation}
where we introduced the notation $Q_\pm=\epsilon+\eta_2 q^{\prime 2}_\pm$. Since the main contribution to the above integral with respect to $z$ comes from the small values of $z \ll T$, we can expand $\operatorname{coth}(z / 2 T) \approx 2 T / z$ and complete the integral analytically leading to 
\begin{equation}
S=\frac{\nu^{-2} i\omega}{16} \frac{\pi}{\left(\epsilon+\eta_{2} q_{+}^{\prime 2}\right)\left(\epsilon+\eta_{2} q_{-}^{\prime 2}\right)\left[\epsilon+\frac{\eta_{2}}{2}\left(q_{+}^{\prime 2}+q_{-}^{\prime 2}\right)\right]}.
\end{equation}
As a next step, we evaluate the momentum integral to obtain $\chi_{i,j}(\mathbf{q}, \omega)$, where a factor of $q^{\prime 2}$ arises from the left and right blocks $B_{i,j}(\mathbf{q}^\prime)$. This brings us to the
following expression 
\begin{equation}
S_{ij}=\frac{\nu^{-2} i \pi \omega}{8} \int_{\mathbf{q}}\int_{0}^{\infty} \frac{q^{\prime 3} d q^{\prime}}{2 \pi}  \int \frac{d \hat{q}^\prime}{2 \pi} \frac{\left[\mathbf{c} \times \mathbf{\hat{q}^\prime}\right]_{i}\left[\mathbf{c} \times \mathbf{\hat{q}^\prime}\right]_{j}}{Q_+Q_-(Q_++Q_-)}.
\end{equation}
 Here, we have assumed that the block functions $B_{i,j}(\mathbf{q^\prime})$ are independent of $\mathbf{q}$. Indeed, terms in the expansion of $B_{i,j}(\mathbf{q^\prime})$ that are linear in $\mathbf{q}$ can be shown to vanish identically upon integration. Furthermore, terms quadratic in $\mathbf{q}$ do not appear at zeroth order in $\mathbf{q^\prime}$. While terms quadratic in $\mathbf{q}$ can emerge at first order in $\mathbf{q^\prime}$, we safely neglect them as the resulting contributions are cubic in $(\mathbf{q}, \mathbf{q^\prime})$. Given that
$q_{\pm}^{\prime 2}=q^{\prime 2} + \frac{q^2}{4} \pm q^\prime q \cos \phi$  and $q_{+}^{\prime 2} + q_{-}^{\prime 2} = 2 q^{\prime 2} + \frac{q^2}{2}$,
where $\phi$ is the angle between $\mathbf{q^\prime}$ and $\mathbf{q}$, we can evaluate the angular integration first using Eq. \eqref{eq:q-integral-phi}.
The radial part of the $q'$ integral then follows 
\begin{align}
\int_{0}^{\infty} \frac{q^{\prime 3} d q^{\prime}}{2 \pi} &\frac{1}{Q_+Q_-(Q_++Q_-)}=\nonumber \\ 
&\left(\frac{1}{16\pi \eta_2^2 \epsilon}\right)\frac{\tan^{-1}\left[\frac{x \cos(\phi)}{\sqrt{1+x^2 \sin^2(\phi)}}\right] }{x \cos(\phi) \sqrt{1+x^2 \sin^2(\phi)}},
\end{align}
with the notation $x=\frac{\sqrt{\eta_{2}} q}{2\sqrt{\epsilon}}$. Lastly, to average this expression over the relative angle $\phi$ between $\mathbf{q}$ and $\mathbf{q^\prime}$ we use the integral identity 
\begin{equation}
\int_0^{2\pi}\frac{\tan^{-1}\left[\frac{x \cos(\phi)}{\sqrt{1+x^2 \sin^2(\phi)}}\right] }{x \cos(\phi) \sqrt{1+x^2 \sin^2(\phi)}}\frac{d \phi}{2 \pi} = \frac{\sinh^{-1}(x)}{x\sqrt{1+x^2}}.
\end{equation}
Putting all these pieces together we arrive at the fluctuation-driven correction to the dynamical spin structure factor 
\begin{equation}
\operatorname{Im} \delta\chi_{ij} \left({q,\omega}\right) =\left(\delta_{ij}-c_i c_j\right) \frac{\omega}{16 \epsilon v_F^2} \frac{\sinh^{-1}(x)}{x\sqrt{1+x^2}} \left(\frac{\alpha_R}{v_F}\right)^2 \mathcal{F}(\varkappa,\varrho),
\end{equation}
Finally, to obtain the NMR relaxation rate in Eq. \eqref{eq:K} it remains to perform the last $q$ integral. It needs a regularization at the ultraviolet limit similar to the case of spin susceptibility. 
With the logarithmic accuracy in  $\epsilon \ll 1$ we obtain  
\begin{equation}
\frac{1}{4 \epsilon \eta_{2}^2} \int_0^{\frac{1}{\xi}} \frac{q d q}{2 \pi} \frac{\sinh^{-1}(x)}{x\sqrt{1+x^2}}\approx \frac{1}{16\pi \eta_{2}^3} \ln ^2 \left(\frac{1}{\epsilon}\right).
\end{equation}
The final expression for the fluctuation-driven correction to the NMR relaxation rate arising from the AL process, normalized by the Korringa value and obtained at an arbitrary elastic scattering time and arbitrary SOC strength in units of $T_c$, is given by:
\begin{equation}
\frac{\delta K}{K}=\frac{1}{64m\varepsilon_F\eta_2}\left(\frac{\alpha_R}{v_F}\right)^2 \ln ^2 \left(\frac{1}{\epsilon}\right)\mathcal{F}\left(\varkappa,\varrho\right),
\end{equation}
The limiting cases for the clean limit, when $\tau\gg\max\left(1/T_c,1/\alpha_R p_F\right)$, the disordered limit, and the entire crossover from weak to strong SOC in a parameter $\alpha_Rp_F/\pi T$, can be easily extracted from the asymptotic expressions given in Eqs. \eqref{eq:eta2} and \eqref{eq:spin-sus-clean}--\eqref{eq:spin-sus-dirty-weakSOC}. 

We compare this result to the earlier calculations reported in Ref. \cite{Randeria:94}. It was found that the combined effect of the regular part of the MT diagram and DOS depletion leads to the negative corrections to the NMR rate
that is logarithmic $\delta K/K\simeq -(T_c/\varepsilon_F)\ln(1/\epsilon)$ in both ballistic limit, $T_c\tau\gg1$, and $\delta K/K\simeq -(1/\varepsilon_F\tau)\ln(1/\epsilon)$ in the diffusive limit $T_c \tau\ll1$.     
These contributions compete with the anomalous MT term which is positive, however strongly sensitive to the dephasing effect as controlled by a parameter $T\tau_\phi$, where $\tau_\phi$ is the dephasing time.  
For weak dephasing, $T\tau_\phi\gg1$, anomalous MT term provides the strongest divergence and scales parametrically as follows $\delta K/K\simeq (1/\varepsilon_F\tau)\ln(\epsilon T\tau_\phi)/\epsilon$ so it dominates the response. It is suppressed however in the opposite limit, $T\tau_\phi\ll1$, and behaves as follows $\delta K/K\simeq (1/\varepsilon_F\tau)(T\tau_\phi)\ln\frac{1}{\epsilon T\tau_\phi}$, which is smaller than the regular MT+DOS contribution in the same parameter regime. Therefore, the fluctuational MEE contribution that we find from the AL process is at play in systems where dephasing suppresses the anomalous MT term. In conclusion, we note that although the applicability of our model requires the small parameter $\alpha_R/v_F$, the term we find is more singular in $\epsilon$ than both the regular MT and DOS contributions.

%$$$$$$$$$$$$$$$$$$$$$$$$$$$$$$$$$$$$$$$$$$$$$$$$$$$$$$$$$$$$$$$$$$$$$
%$$$$$$$$$$$$$$$$$$$$$$$$$$$$$$$$$$$$$$$$$$$$$$$$$$$$$$$$$$$$$$$$$$$$$
%$$$$$$$$$$$$$$$$$$$$$$$$$$$$$$$$$$$$$$$$$$$$$$$$$$$$$$$$$$$$$$$$$$$$$
\section*{Acknowledgements}

The work of J. H. was supported by Ames National Laboratory and Iowa State University funds.
The work of D. S. was supported in part by the National Science Foundation (NSF) Quantum Leap Challenge Institute for Hybrid Quantum Architectures and Networks Grant No. OMA-2016136. 
M. D. acknowledges the support of the NSF grant No. DMR-2400484. The work of A. L. was supported by NSF Grant No. DMR-2452658 and H. I. Romnes Faculty Fellowship provided by the University of Wisconsin-Madison Office of the Vice Chancellor for Research and Graduate Education with funding from the Wisconsin Alumni Research Foundation. This paper was written during the workshop program "Emerging New Phases in Quantum Materials: The Disordered, the Strange and the Topological" with the participation of M. D. and A. L. at the Aspen Center for Physics, which is supported by National Science Foundation grant PHY-2210452. 

%$$$$$$$$$$$$$$$$$$$$$$$$$$$$$$$$$$$$$$$$$$$$$$$$$$$$$$$$$$$$$$$$$$$$$
%$$$$$$$$$$$$$$$$$$$$$$$$$$$$$$$$$$$$$$$$$$$$$$$$$$$$$$$$$$$$$$$$$$$$$
%$$$$$$$$$$$$$$$$$$$$$$$$$$$$$$$$$$$$$$$$$$$$$$$$$$$$$$$$$$$$$$$$$$$$$
\appendix

%%%%%%%%%%%%%%%%%%%%%%%%%%%%%%%%%%%%
\begin{figure}[h]
\centering
\includegraphics[width=0.35\textwidth]{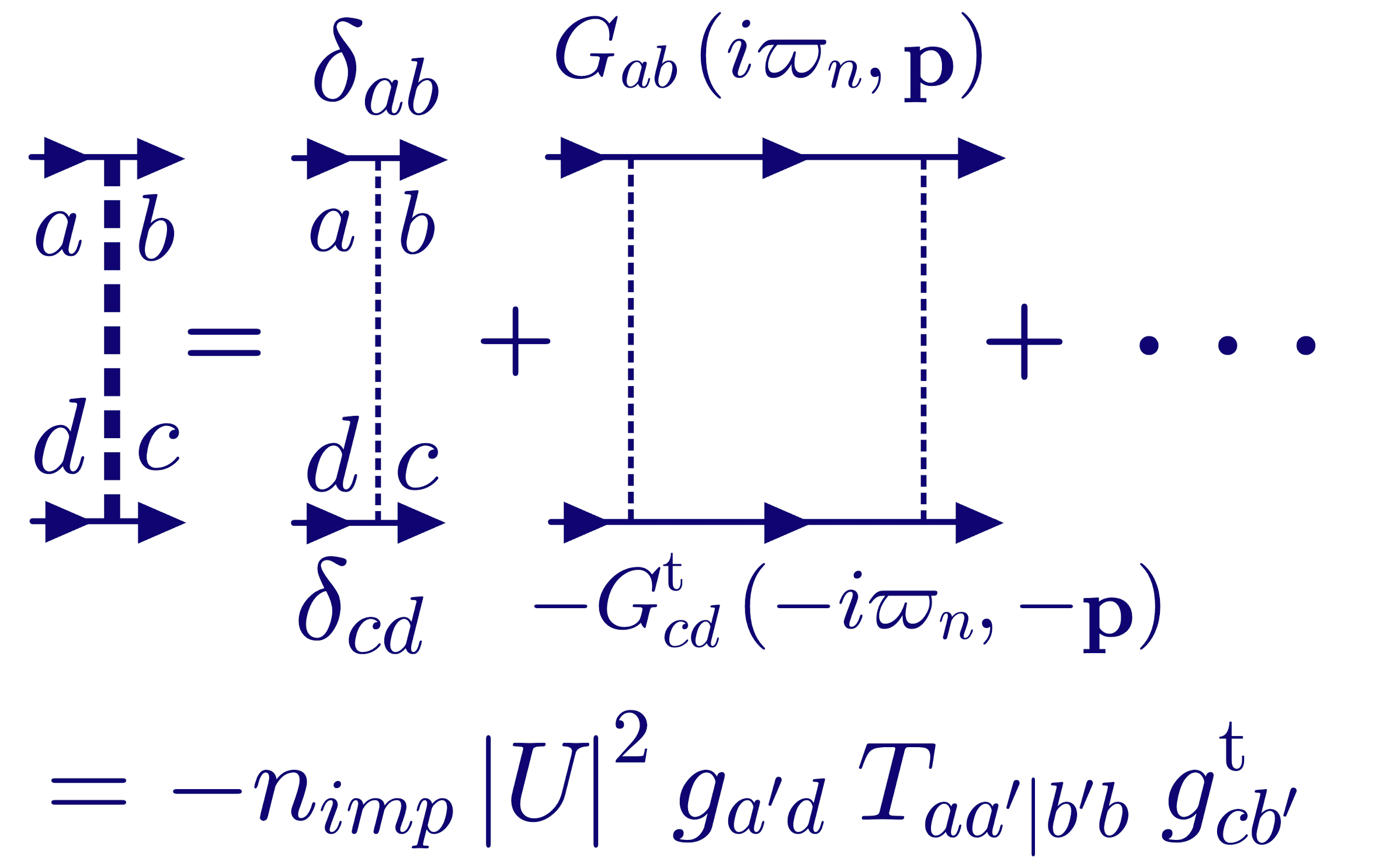} 
\caption{Diagrammatic representation of the $T$-matrix equation for impurity scattering in the Cooper channel.}  \label{fig:Tmatrix}
\end{figure}
%%%%%%%%%%%%%%%%%%%%%%%%%%%%%%%%%%%%

\section{Evaluation of the block \texorpdfstring{$B_{i}(\mathbf{q}^\prime)$}{Bi(q)}}\label{AppA}

In this appendix, we compute the explicit expression for the block $B_{i}(\mathbf{q}^\prime)$ by evaluating the diagrammatic expansion shown in Fig. \ref{fig:Block}. First, we briefly introduce the core components of the diagrammatic technique established in Ref. \cite{Edelstein:21}.

The clockwise-directed fermion lines represent the standard electronic Green's function,
\begin{equation}
G_{ab}\left(i \varpi_n, \mathbf{p}\right) = \sum_{\lambda= \pm} \Pi_{ab}^{(\lambda)}(\mathbf{p}) G_{(\lambda)}\left(i \varpi_n, p\right),
\end{equation}
where $a$ and $b$ denote spin indices. The renormalized imaginary frequency is defined as $i \varpi_n = i \omega_n s\left(\omega_n\right)$, where $s\left(\omega_n\right) = 1 + \left(2 \tau\left|\omega_n\right|\right)^{-1}$ incorporates the elastic scattering time $\tau$, and $\omega_n = (2n+1)\pi T$ represents the discrete fermionic Matsubara frequencies. The operator
\begin{equation}
\Pi^{(\pm)}_{ab}(\mathbf{p}) = \frac{1}{2}\left(\delta_{ab} \pm \frac{[\mathbf{p} \times \mathbf{c}] \cdot \bm{\sigma}_{ab}}{|\mathbf{p} \times \mathbf{c}|}\right),
\end{equation}
is the projection operator onto states with a definite helicity $\lambda = \pm$, where the corresponding single-particle Green's function is given by
\begin{equation}\label{eq:GF_branch}
G_{(\lambda)}\left(i \varpi_n, p\right) = \left[i \varpi_n - \xi_{(\lambda)}(p)\right]^{-1},
\end{equation}
with the helicity-dependent dispersion relation $\xi_{(\pm)}(p) = \frac{p^2}{2 m} -\mu\pm\alpha_{R} p$.

Conversely, the counterclockwise-directed fermion lines represent the transposed Green's function, $-G_{ab}^{\mathrm{t}}\left(-i \varpi_n, -\mathbf{p}\right)$. For algebraic convenience, we define the reversed Green's function, $G_{ab}^{(\mathrm{r})}\left(i \varpi_n, \mathbf{p}\right)$, via the relation
\begin{equation}
-G_{ab}^{\mathrm{t}}\left(-i \varpi_n, -\mathbf{p}\right) = g_{ac}^{\mathrm{t}} G_{cd}^{(\mathrm{r})}\left(i \varpi_n, \mathbf{p}\right) g_{db},
\end{equation}
which expands in the helicity basis as
\begin{align}
&G_{ab}^{(\mathrm{r})}\left(i \varpi_n, \mathbf{p}\right) = \sum_{\lambda= \pm} \Pi_{ab}^{(\lambda)}(\mathbf{p}) G_{(\lambda)}^{(\mathrm{r})}\left(i \varpi_n, p\right), \nonumber \\ 
&G_{(\lambda)}^{(\mathrm{r})}\left(i \varpi_n, p\right) = \left[i \varpi_n + \xi_{(\lambda)}(p)\right]^{-1}.
\end{align}

The small hollow circles, which stem from the $\mathbf{q}^\prime$-expansion of the Green's functions, represent the insertion $-\mathbf{q}^\prime \cdot \mathbf{v}^{\mathrm{t}}(-\mathbf{p})$ when they occur along a counterclockwise fermion line. Here, the velocity operator is given by
\begin{equation}
\mathbf{v}_{ab}(\mathbf{p}) = i\left[H^{(0)}_{ab}(\mathbf{p}), \mathbf{r}\right] = \frac{\mathbf{p}}{m} \delta_{ab} + \alpha_{R}[\mathbf{c} \times \bm{\sigma}]_{ab}.
\end{equation}

The impurity ladders are represented by thick dashed lines, defined diagrammatically in Fig. \ref{fig:Tmatrix}. The corresponding vertex scattering matrix $\widehat{T}\left(\omega_n\right)$ is expressed as
\begin{subequations}
\begin{align}\label{eq:Tmatrix}
T_{ad \mid cb}\left(\omega_n\right) = \frac{1}{2}\left[\frac{\delta_{ad} \delta_{cb}}{1-w\left(\omega_n\right)} + \frac{(\mathbf{c} \cdot \bm{\sigma})_{ad}(\mathbf{c} \cdot \bm{\sigma})_{cb}}{1-u\left(\omega_n\right)}\right. \nonumber \\ 
\left. + \frac{[\mathbf{c} \times \bm{\sigma}]_{ad}^i[\mathbf{c} \times \bm{\sigma}]_{cb}^i}{1-v\left(\omega_n\right)}\right],
\end{align}
where the dimensionless impurity correlation functions $w\left(\omega_n\right)$, $u\left(\omega_n\right)$, and $v\left(\omega_n\right)$ take the forms
\begin{align}
& w\left(\omega_n\right) = \frac{1}{2 \tau\left|\omega_n\right| + 1}, \\ 
& u\left(\omega_n\right) = \frac{2 \tau\left|\omega_n\right| + 1}{\left(2 \tau\left|\omega_n\right| + 1\right)^2 + 4\varkappa^2},\\ 
& v\left(\omega_n\right) = \frac{1}{2}\left[w\left(\omega_n\right) + u\left(\omega_n\right)\right].
\end{align}
\end{subequations}

Finally, the impurity-renormalized vertices are denoted by small solid circles and are assigned the value $\hat{g}s(\omega_n)$ or $\hat{g}^{\mathrm{t}}s(\omega_n)$, depending on whether they correspond to the creation or annihilation of a fluctuating pair.

%%%%%%%%%%%%%%%%%%%%%%%%%%%%%%%%%%%%
\begin{figure}
\centering
\includegraphics[width=0.35\textwidth]{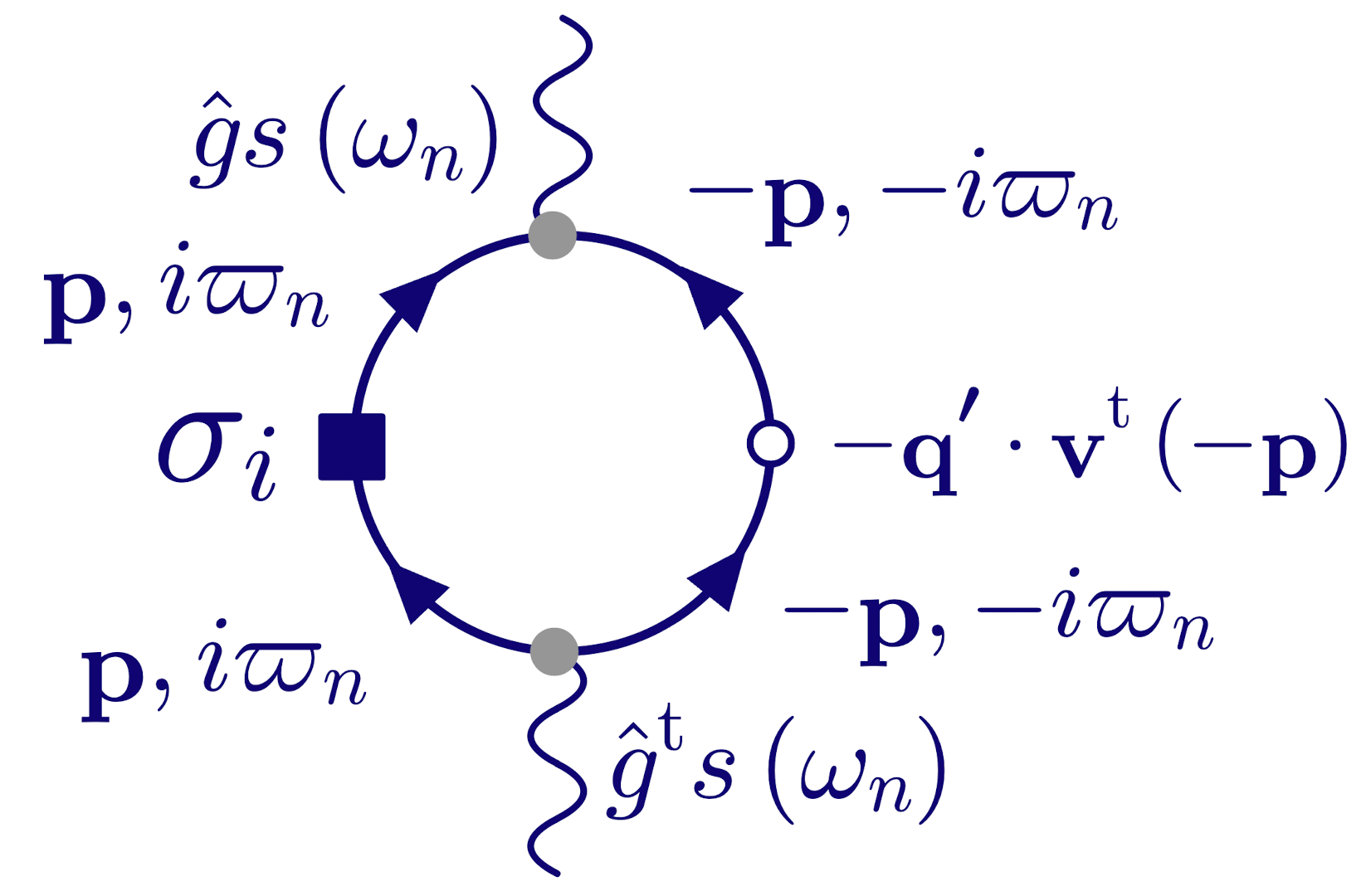} 
\caption{The detailed form of the first diagram responsible for $B_i(\mathbf{q^\prime})$ shown on the last line in Fig. \ref{fig:Block}.} \label{fig:block1}
\end{figure}
%%%%%%%%%%%%%%%%%%%%%%%%%%%%%%%%%%%%

%$$$$$$$$$$$$$$$$$$$$$$$$$$$$$$$$$$$$$$$$$$$$$$$$$$$$$$$$$$$$$$$$$$$$$
%$$$$$$$$$$$$$$$$$$$$$$$$$$$$$$$$$$$$$$$$$$$$$$$$$$$$$$$$$$$$$$$$$$$$$
%$$$$$$$$$$$$$$$$$$$$$$$$$$$$$$$$$$$$$$$$$$$$$$$$$$$$$$$$$$$$$$$$$$$$$
\section{Evaluation of the diagram without an impurity ladder}\label{AppA1}

In this appendix we evaluate the diagram without an impurity ladder shown in Fig. \ref{fig:block1}. This diagram is evaluated as follows:
\begin{align}
& B_{i,1}(\mathbf{q^\prime})=T \sum_{\omega_n} s^2\left(\omega_n\right) q^\prime_j \nonumber \\ 
&\times
\int_{\mathbf{p}} \sum_{\lambda \lambda^{\prime}} \operatorname{Tr}\left\{\mathbf{\sigma}_i \Pi^{(\lambda)}(\mathbf{p}) v_j(\mathbf{p}) \Pi^{(\lambda^{\prime})}(\mathbf{p})\right\}G_{(\lambda)} G_{(\lambda)}^{\text {(r)}} G_{(\lambda^{\prime})} G_{(\lambda^{\prime})}^{(\text{r})},
\end{align}
where $\int_{\mathbf{p}}=2\pi\nu \int \frac{d \xi}{2 \pi} \int \frac{d \hat{p}}{2 \pi}$ and all the $G$'s and $G^{\mathrm{(r)}}$'s have the same argument $\left(i \varpi_n, \mathbf{p}\right)$. Here, the angular integral has the following expression,
\begin{align}
&\int \frac{d \hat{p}}{2 \pi} \operatorname{Tr}\left\{\mathbf{\sigma}_i \Pi^{(\lambda)}(\mathbf{p}) v_j(\mathbf{p}) \Pi^{(\lambda^{\prime})}(\mathbf{p})\right\} \nonumber \\ 
&=\frac{1}{2} \varepsilon_{i j k} c_k\left[\frac{1}{2}(\lambda+\lambda^{\prime}) \frac{p}{m}+\alpha_R\right]. 
\end{align}
Using this result the vertex function can be brought to the form 
\begin{align}
&B_{i,1}(\mathbf{q^\prime}) = -\pi \nu \left[\mathbf{c} \times \mathbf{q^\prime}\right]_{i} T \sum_{\omega_n} s^2\left(\omega_n\right) \nonumber \\ 
&\sum_{\lambda=\pm1} \int \frac{d \xi}{2 \pi} \biggl\{\left(\lambda \frac{p}{m}+\alpha_R\right)G_{(\lambda)}^2 \left[G_{(\lambda)}^{\text {(r)}}\right]^2+\alpha_R G_{(\lambda)} G_{(\lambda)}^{\text {(r)}} G_{(-\lambda)} G_{(-\lambda)}^{(\text{r})}\biggr\},
\end{align}
This can now be further simplified using the following integrals:
\begin{subequations}
\begin{align}
&\int \frac{d \xi}{2 \pi} G_{(\lambda)}^2 \left[G_{(\lambda)}^{(r)}\right]^2=(1-\lambda \delta) \frac{2}{\left[2|\varpi_n|\right]^3}, \\ 
&\int \frac{d \xi}{2 \pi} \left(\frac{p}{m}\right) G_{(\lambda)}^2 \left[G_{(\lambda)}^{(r)}\right]^2=(1-2\lambda \delta) \frac{2}{\left[2|\varpi_n|\right]^3}, \\
&\int \frac{d \xi}{2 \pi} G_{(\lambda)} G_{(\lambda)}^{(r)} G_{(-\lambda)} G_{(-\lambda)}^{(r)}=\frac{1}{\left|\varpi_n\right|}  \frac{1}{\left[2\left|\varpi_n\right|\right]^2+(2\alpha_R p_F)^2}.
\end{align}
\end{subequations}
Finally, we get 
\begin{equation}
B_{i,1}(\mathbf{q^\prime}) =  T \sum_{\omega_n} s^2\left(\omega_n\right)
\frac{4\pi \nu \alpha_R \left[\mathbf{c} \times \mathbf{q^\prime}\right]_{i}\left(2 \alpha_R p_F\right)^2}{\left(2\left|\varpi_n\right|\right)^3\left[\left(2\varpi_n\right)^2+\left(2 \alpha_R p_F\right)^2\right]}.
\end{equation}

%@@@@@@@@@@@@@@@@@@@@@@@@@@@@@@@@@@@@@@@@@@@@@@@@@@@@@@@@@@@@@@@@@@@@@@@@@@@@@@@@@@@@@
\begin{figure}
\includegraphics[width=0.45\textwidth]{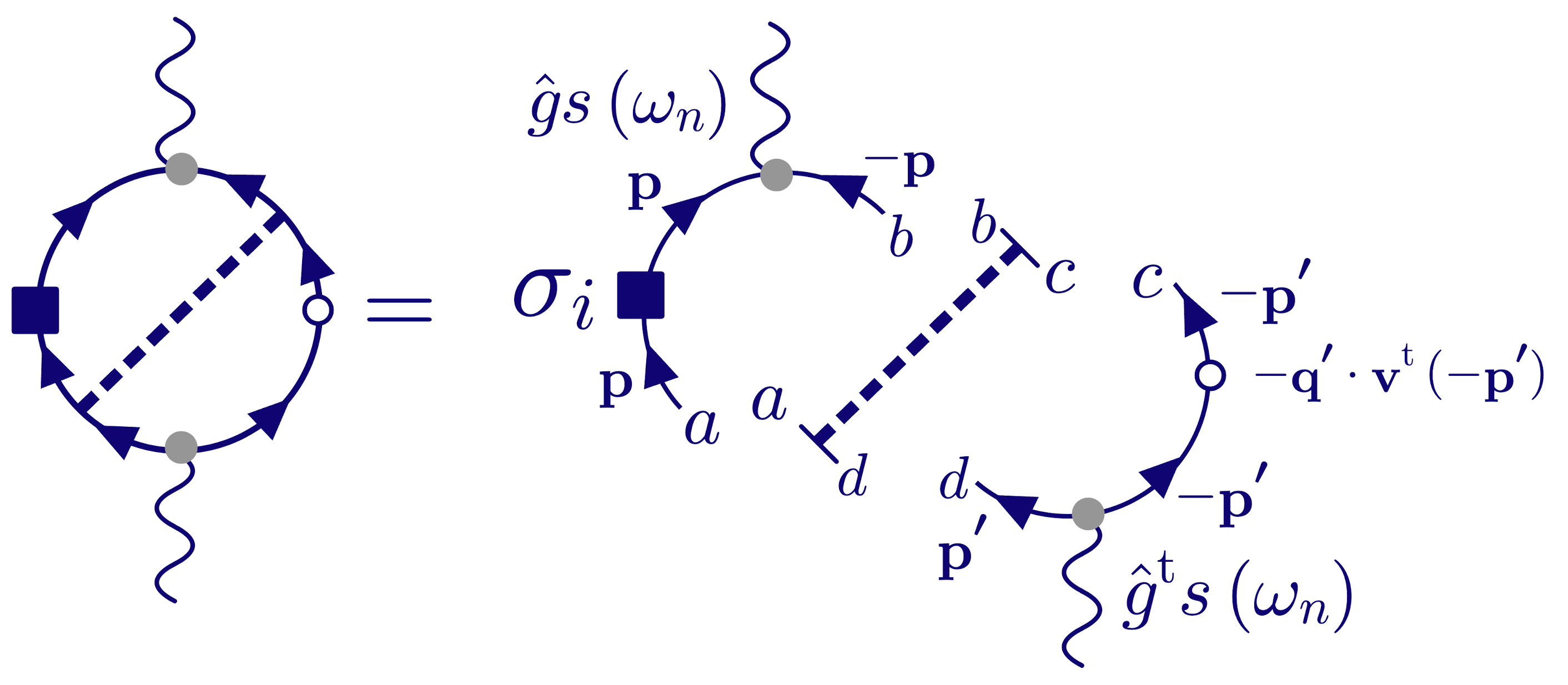} 
\centering
\caption{The second diagram contributing to the vertex block $B_i(\mathbf{q^\prime})$ is shown in the last line of Fig. \ref{fig:Block}. The split form of this diagram is depicted on the right.}  
\label{fig:block2}
\end{figure}
%@@@@@@@@@@@@@@@@@@@@@@@@@@@@@@@@@@@@@@@@@@@@@@@@@@@@@@@@@@@@@@@@@@@@@@@@@@@@@@@@@@@@@

%$$$$$$$$$$$$$$$$$$$$$$$$$$$$$$$$$$$$$$$$$$$$$$$$$$$$$$$$$$$$$$$$$$$$$
%$$$$$$$$$$$$$$$$$$$$$$$$$$$$$$$$$$$$$$$$$$$$$$$$$$$$$$$$$$$$$$$$$$$$$
%$$$$$$$$$$$$$$$$$$$$$$$$$$$$$$$$$$$$$$$$$$$$$$$$$$$$$$$$$$$$$$$$$$$$$
\section{Evaluation of diagrams with an impurity ladder}\label{AppA2}

We complete the calculation by evaluating one diagram with an impurity ladder shown in Fig. \ref{fig:block2}. The analytical structure of this diagram can be presented as follows 
\begin{align}
B_{i,2}(\mathbf{q^\prime})= &\left(-n_{\text {imp }}|U|^2\right) q^\prime_j \nonumber \\ \times &T \sum_{\omega_n} s^2\left(\omega_n\right) \operatorname{Tr}\left\{L\left(\omega_n\right) \circ T\left(\omega_n\right) \circ R\left(\omega_n\right)\right\},
\end{align}
where $T\left(\omega_n\right)$ is given by Eq. (\ref{eq:Tmatrix}), $L\left(\omega_n\right)$ and $R\left(\omega_n\right)$ denote the left and right fragments of the diagram in its split form (see Fig. \ref{fig:block2}): 
\begin{align}
L_{ab}&=\int_{\mathbf{p}}\left\{G\left(i \varpi_n, \mathbf{p}\right) \mathbf{\sigma}_i G\left(i \varpi_n, \mathbf{p}\right) G^{(\mathrm{r})}\left(i\varpi_n, \mathbf{p}\right)\right\}_{ab}\nonumber \\ 
&=\int_{\mathbf{p}} \sum_{\lambda \lambda^{\prime}} G_{(\lambda^{\prime})} G_{(\lambda)} G_{(\lambda)}^{(\mathrm{r})}\left\{\Pi^{(\lambda^{\prime})}\sigma_i \Pi^{(\lambda)}\right\}_{ab},
\end{align}
where arguments of the Green's functions and the projection operators are suppressed for brevity. 
The angular integral gives
\begin{equation}
\int \frac{d \hat{p}}{2 \pi}\left[\Pi^{(\lambda^{\prime})} \sigma_i \Pi^{(\lambda)}\right]_{ab}=\frac{1}{4}\left[\sigma_i-\lambda \lambda^{\prime} c_i(\mathbf{c} \cdot \bm{\sigma})\right]_{ab}.
\end{equation}
Using this expression $L_{ab}$ takes the following form
\begin{equation}\label{eq:Lb}
\begin{aligned}
L_{ab}=&2\pi\nu\sum_{\lambda= \pm 1} \left\{\frac{1}{4}\left[\sigma_i-c_i(\mathbf{c} \cdot \bm{\sigma})\right]_{ab} \int \frac{d \xi}{2 \pi} G_{(\lambda)}^2 G_{(\lambda)}^{(r)}\right. \\
& \left.+\frac{1}{4}\left[\sigma_i+c_i(\mathbf{c} \cdot \bm{\sigma})\right]_{ab} \int \frac{d \xi}{2 \pi} G_{(\lambda)} G_{(\lambda)}^{(r)} G_{(-\lambda)}\right\}.
\end{aligned}
\end{equation}
Note that the last term in Eq. (\ref{eq:Lb}) is not diagonal in the helical indices, i.e. $\lambda=-\lambda^{\prime}$ terms also contribute to $L_{ab}$. This can be simplified using the following integrals:
\begin{subequations}
\begin{equation}
\int \frac{d \xi}{2 \pi} G_{(\lambda)}^2 G_{(\lambda)}^{(r)}=(1-\lambda \delta) \frac{i \operatorname{sgn}\left(\omega_n\right)}{\left[2 \varpi_n\right]^2},
\end{equation}
\begin{equation}\label{eq:off_diag21}
\int \frac{d \xi}{2 \pi} G_{(\lambda)} G_{(\lambda)}^{(r)} G_{(-\lambda)}=\frac{i \operatorname{sgn}\left(\omega_n\right)}{\left(2\varpi_n\right)^2-4 i \lambda\left|\varpi_n\right| \alpha_R p_F}.
\end{equation}
\end{subequations}
We finally get for $L_{ab}$:
\begin{equation}\label{eq:left}
L_{ab}=\pi\nu i \operatorname{sgn}\omega_n\left[\frac{\sigma_i-c_i(\mathbf{c} \cdot \bm{\sigma})}{\left(2\varpi_n\right)^2}+\frac{\sigma_i+c_i(\mathbf{c} \cdot \bm{\sigma})}{\left(2\varpi_n\right)^2+(2\alpha_Rp_F)^2}\right]_{ab},
\end{equation}

We next proceed to evaluate the right fragment of the diagram,
\begin{equation}
\begin{aligned}
R_{cd}&=\int_{\mathbf{p}}\left\{G^{(r)}\left(i\varpi_n, \mathbf{p}\right) v^j(\mathbf{p}) G^{(r)}\left(i\varpi_n, \mathbf{p}\right) G\left(i\varpi_n, \mathbf{p}\right)\right\}_{cd}\\
&=\int_{\mathbf{p}} \sum_{\lambda \lambda^{\prime}} G_{(\lambda^{\prime})}^{(r)} G_{(\lambda)}^{(r)} G_{(\lambda)}\left\{\Pi^{(\lambda^{\prime})} v^j(\mathbf{p}) \Pi^{(\lambda)}\right\}_{cd},
\end{aligned}
\end{equation}
where the angular integral is given by
\begin{equation}
\begin{aligned}
&\int \frac{d \hat{p}}{2 \pi}\left[\Pi^{(\lambda^{\prime})} v^j(\mathbf{p}) \Pi^{(\lambda)}\right]_{cd}=\frac{1}{4}[\mathbf{c} \times \bm{\sigma}]^j_{cd}\left[\alpha_R+\frac{1}{2}(\lambda+\lambda^{\prime})\left(\frac{p}{m}\right)\right].
\end{aligned}
\end{equation}
Using this expression $R_{ab}$ reduces to the form
\begin{align}\label{eq:Rb}
&R_{cd}=\pi\nu \frac{1}{2}[\mathbf{c} \times \bm{\sigma}]^j_{cd}\nonumber \\ 
&\sum_{\lambda= \pm 1} \int \frac{d \xi}{2 \pi}\biggl\{\left[\alpha_R +\lambda\left(\frac{p}{m}\right)\right] G_{(\lambda)} \left[G_{(\lambda)}^{(r)}\right]^2
+\alpha_R G_{(\lambda)} G_{(\lambda)}^{(r)} G_{(-\lambda)}^{(r)}\biggr\},
\end{align}
This can be simplified using the following integrals:
\begin{subequations}
\begin{equation}
\int \frac{d \xi}{2 \pi} G_{(\lambda)} \left[G_{(\lambda)}^{(r)}\right]^2=(1-\lambda \delta) \frac{i \operatorname{sgn}\left(\omega_n\right)}{\left[2 \varpi_n\right]^2},
\end{equation}
\begin{equation}\label{eq:off_diag21r}
\int \frac{d \xi}{2 \pi} G_{(\lambda)} G_{(\lambda)}^{(r)} G_{(-\lambda)}^{(r)}=\frac{i \operatorname{sgn}\left(\omega_n\right)}{\left(2\varpi_n\right)^2+4 i \lambda\left|\varpi_n\right| \alpha_R p_F}.
\end{equation}
\end{subequations}
Putting the pieces together, we finally get for $R_{cd}$:
\begin{equation}\label{eq:right}
R_{cd}=-\pi\nu i \operatorname{sgn}\left(\omega_n\right)\left[\frac{\alpha_R(\mathbf{c} \times \bm{\sigma})_j (2\alpha_Rp_F)^2}{\left[2\varpi_n\right]^2\left[\left(2 \varpi_n\right)^2+(2\alpha_Rp_F)^2\right]}\right]_{cd}.
\end{equation}
From the form of Eq. (\ref{eq:Tmatrix}), (\ref{eq:left}) and (\ref{eq:right}), it can be seen that $T\left(\omega_n\right)$ connects $L\left(\omega_n\right)$ and $R\left(\omega_n\right)$ with its last term
\begin{equation}
\frac{1}{2} \frac{(\mathbf{c} \times \bm{\sigma})_{ba}^l(\mathbf{c} \times \bm{\sigma})_{dc}^l}{1-v\left(\omega_n\right)}.
\end{equation}
We thus find for $B_{i,2}(\mathbf{q^\prime})$:
\begin{align}\label{eq:Q11b}
B_{i,2}(\mathbf{q^\prime})=&-\frac{q^\prime_j}{4\pi\nu\tau} T \sum_{\omega_n} \frac{s^2\left(\omega_n\right)}{1-v\left(\omega_n\right)}\nonumber \\ 
&\times \operatorname{Tr}\left[L\left(\omega_n\right)[\mathbf{c} \times \bm{\sigma}]^l\right] \operatorname{Tr}\left[R\left(\omega_n\right)[\mathbf{c} \times \bm{\sigma}]^l\right]
\end{align}
Computing the remaining traces gives the result 
\begin{align}
B_{i,2}(\mathbf{q^\prime})&=2\pi\nu\alpha_R\left[\mathbf{c} \times \mathbf{q^\prime}\right]_{i} 
\nonumber \\ 
&\times T \sum_{\omega_n>0} \frac{1}{\omega_n^2} \frac{\tau \varkappa^2\left[\left(2 \tau\omega_n+1\right)^2+2\varkappa^2\right]}{\left(2 \tau\omega_n+1\right)\left[\left(2 \tau\omega_n+1\right)^2+4\varkappa^2\right]} 
\nonumber \\
& \times \frac{1}{\tau\omega_n\left(2 \tau\omega_n+1\right)^2+\varkappa^2\left(4 \tau\omega_n+1\right)}.
\end{align}
The remaining diagram with the impurity ladder evaluates to the same expression as above. We thus arrive at Eq. (\ref{eq:block}) after adding the contributions from all three diagrams:
\begin{equation}
    B_{i}(\mathbf{q^\prime})=B_{i,1}(\mathbf{q^\prime})+2 B_{i,2}(\mathbf{q^\prime}).
\end{equation}
This provides the explicit forms for all the building blocks of the diagrammatic technique used in the main text.

%$$$$$$$$$$$$$$$$$$$$$$$$$$$$$$$$$$$$$$$$$$$$$$$$$$$$$$$$$$$$$$$$$$$$$
%$$$$$$$$$$$$$$$$$$$$$$$$$$$$$$$$$$$$$$$$$$$$$$$$$$$$$$$$$$$$$$$$$$$$$
%$$$$$$$$$$$$$$$$$$$$$$$$$$$$$$$$$$$$$$$$$$$$$$$$$$$$$$$$$$$$$$$$$$$$$
\bibliography{biblio-Spin-NMR-Fluctuations}

\begin{thebibliography}{58}%
\makeatletter
\providecommand \@ifxundefined [1]{%
 \@ifx{#1\undefined}
}%
\providecommand \@ifnum [1]{%
 \ifnum #1\expandafter \@firstoftwo
 \else \expandafter \@secondoftwo
 \fi
}%
\providecommand \@ifx [1]{%
 \ifx #1\expandafter \@firstoftwo
 \else \expandafter \@secondoftwo
 \fi
}%
\providecommand \natexlab [1]{#1}%
\providecommand \enquote  [1]{``#1''}%
\providecommand \bibnamefont  [1]{#1}%
\providecommand \bibfnamefont [1]{#1}%
\providecommand \citenamefont [1]{#1}%
\providecommand \href@noop [0]{\@secondoftwo}%
\providecommand \href [0]{\begingroup \@sanitize@url \@href}%
\providecommand \@href[1]{\@@startlink{#1}\@@href}%
\providecommand \@@href[1]{\endgroup#1\@@endlink}%
\providecommand \@sanitize@url [0]{\catcode `\\12\catcode `\$12\catcode
  `\&12\catcode `\#12\catcode `\^12\catcode `\_12\catcode `\%12\relax}%
\providecommand \@@startlink[1]{}%
\providecommand \@@endlink[0]{}%
\providecommand \url  [0]{\begingroup\@sanitize@url \@url }%
\providecommand \@url [1]{\endgroup\@href {#1}{\urlprefix }}%
\providecommand \urlprefix  [0]{URL }%
\providecommand \Eprint [0]{\href }%
\providecommand \doibase [0]{http://dx.doi.org/}%
\providecommand \selectlanguage [0]{\@gobble}%
\providecommand \bibinfo  [0]{\@secondoftwo}%
\providecommand \bibfield  [0]{\@secondoftwo}%
\providecommand \translation [1]{[#1]}%
\providecommand \BibitemOpen [0]{}%
\providecommand \bibitemStop [0]{}%
\providecommand \bibitemNoStop [0]{.\EOS\space}%
\providecommand \EOS [0]{\spacefactor3000\relax}%
\providecommand \BibitemShut  [1]{\csname bibitem#1\endcsname}%
\let\auto@bib@innerbib\@empty
%</preamble>
\bibitem [{\citenamefont {Mineev}\ and\ \citenamefont
  {Samokhin}(1999)}]{Mineev:99}%
  \BibitemOpen
  \bibfield  {author} {\bibinfo {author} {\bibfnamefont {Vladimir~P.}\
  \bibnamefont {Mineev}}\ and\ \bibinfo {author} {\bibfnamefont {Kirill~V.}\
  \bibnamefont {Samokhin}},\ }\href@noop {} {\emph {\bibinfo {title}
  {Introduction to Unconventional Superconductivity}}}\ (\bibinfo  {publisher}
  {CRC Press},\ \bibinfo {year} {1999})\BibitemShut {NoStop}%
\bibitem [{\citenamefont {Bauer}\ and\ \citenamefont
  {Sigrist}(2012)}]{Bauer:12}%
  \BibitemOpen
  \bibinfo {editor} {\bibfnamefont {Ernst}\ \bibnamefont {Bauer}}\ and\
  \bibinfo {editor} {\bibfnamefont {Manfred}\ \bibnamefont {Sigrist}},\ eds.,\
  \href {\doibase 10.1007/978-3-642-24624-1} {\emph {\bibinfo {title}
  {Non-Centrosymmetric Superconductors: Introduction and Overview}}},\ \bibinfo
  {series} {Lecture Notes in Physics}, Vol.\ \bibinfo {volume} {847}\ (\bibinfo
   {publisher} {Springer Berlin Heidelberg},\ \bibinfo {year}
  {2012})\BibitemShut {NoStop}%
\bibitem [{\citenamefont {Lu}\ \emph {et~al.}(2015)\citenamefont {Lu},
  \citenamefont {Zheliuk}, \citenamefont {Leermakers}, \citenamefont {Yuan},
  \citenamefont {Zeitler}, \citenamefont {Law},\ and\ \citenamefont
  {Ye}}]{Lu:15}%
  \BibitemOpen
  \bibfield  {author} {\bibinfo {author} {\bibfnamefont {J.~M.}\ \bibnamefont
  {Lu}}, \bibinfo {author} {\bibfnamefont {O.}~\bibnamefont {Zheliuk}},
  \bibinfo {author} {\bibfnamefont {I.}~\bibnamefont {Leermakers}}, \bibinfo
  {author} {\bibfnamefont {N.~F.~Q.}\ \bibnamefont {Yuan}}, \bibinfo {author}
  {\bibfnamefont {U.}~\bibnamefont {Zeitler}}, \bibinfo {author} {\bibfnamefont
  {K.~T.}\ \bibnamefont {Law}}, \ and\ \bibinfo {author} {\bibfnamefont
  {J.~T.}\ \bibnamefont {Ye}},\ }\bibfield  {title} {\enquote {\bibinfo {title}
  {Evidence for two-dimensional {Ising} superconductivity in gated
  {M}o{S}$_2$},}\ }\href {\doibase 10.1126/science.aab2277} {\bibfield
  {journal} {\bibinfo  {journal} {Science}\ }\textbf {\bibinfo {volume}
  {350}},\ \bibinfo {pages} {1353--1357} (\bibinfo {year} {2015})}\BibitemShut
  {NoStop}%
\bibitem [{\citenamefont {Kang}\ \emph {et~al.}(2015)\citenamefont {Kang},
  \citenamefont {Zhou}, \citenamefont {Yi}, \citenamefont {Yang}, \citenamefont
  {Guo}, \citenamefont {Shi}, \citenamefont {Zhang}, \citenamefont {Wang},
  \citenamefont {Zhang}, \citenamefont {Jiang}, \citenamefont {Li},
  \citenamefont {Yang}, \citenamefont {Wu}, \citenamefont {Zhang},
  \citenamefont {Sun},\ and\ \citenamefont {Zhao}}]{Kang:15}%
  \BibitemOpen
  \bibfield  {author} {\bibinfo {author} {\bibfnamefont {Defen}\ \bibnamefont
  {Kang}}, \bibinfo {author} {\bibfnamefont {Yazhou}\ \bibnamefont {Zhou}},
  \bibinfo {author} {\bibfnamefont {Wei}\ \bibnamefont {Yi}}, \bibinfo {author}
  {\bibfnamefont {Chongli}\ \bibnamefont {Yang}}, \bibinfo {author}
  {\bibfnamefont {Jing}\ \bibnamefont {Guo}}, \bibinfo {author} {\bibfnamefont
  {Youguo}\ \bibnamefont {Shi}}, \bibinfo {author} {\bibfnamefont {Shan}\
  \bibnamefont {Zhang}}, \bibinfo {author} {\bibfnamefont {Zhe}\ \bibnamefont
  {Wang}}, \bibinfo {author} {\bibfnamefont {Chao}\ \bibnamefont {Zhang}},
  \bibinfo {author} {\bibfnamefont {Sheng}\ \bibnamefont {Jiang}}, \bibinfo
  {author} {\bibfnamefont {Aiguo}\ \bibnamefont {Li}}, \bibinfo {author}
  {\bibfnamefont {Ke}~\bibnamefont {Yang}}, \bibinfo {author} {\bibfnamefont
  {Qi}~\bibnamefont {Wu}}, \bibinfo {author} {\bibfnamefont {Guangming}\
  \bibnamefont {Zhang}}, \bibinfo {author} {\bibfnamefont {Liling}\
  \bibnamefont {Sun}}, \ and\ \bibinfo {author} {\bibfnamefont {Zhongxian}\
  \bibnamefont {Zhao}},\ }\bibfield  {title} {\enquote {\bibinfo {title}
  {Superconductivity emerging from a suppressed large magnetoresistant state in
  tungsten ditelluride},}\ }\href {\doibase 10.1038/ncomms8804} {\bibfield
  {journal} {\bibinfo  {journal} {Nature Communications}\ }\textbf {\bibinfo
  {volume} {6}},\ \bibinfo {pages} {7804} (\bibinfo {year} {2015})}\BibitemShut
  {NoStop}%
\bibitem [{\citenamefont {Ugeda}\ \emph {et~al.}(2016)\citenamefont {Ugeda},
  \citenamefont {Bradley}, \citenamefont {Zhang}, \citenamefont {Onishi},
  \citenamefont {Chen}, \citenamefont {Ruan}, \citenamefont
  {Ojeda-Aristizabal}, \citenamefont {Ryu}, \citenamefont {Edmonds},
  \citenamefont {Tsai}, \citenamefont {Riss}, \citenamefont {Mo}, \citenamefont
  {Lee}, \citenamefont {Zettl}, \citenamefont {Hussain}, \citenamefont {Shen},\
  and\ \citenamefont {Crommie}}]{Ugeda:16}%
  \BibitemOpen
  \bibfield  {author} {\bibinfo {author} {\bibfnamefont {Miguel~M.}\
  \bibnamefont {Ugeda}}, \bibinfo {author} {\bibfnamefont {Aaron~J.}\
  \bibnamefont {Bradley}}, \bibinfo {author} {\bibfnamefont {Yi}~\bibnamefont
  {Zhang}}, \bibinfo {author} {\bibfnamefont {Seita}\ \bibnamefont {Onishi}},
  \bibinfo {author} {\bibfnamefont {Yi}~\bibnamefont {Chen}}, \bibinfo {author}
  {\bibfnamefont {Wei}\ \bibnamefont {Ruan}}, \bibinfo {author} {\bibfnamefont
  {Claudia}\ \bibnamefont {Ojeda-Aristizabal}}, \bibinfo {author}
  {\bibfnamefont {Hyejin}\ \bibnamefont {Ryu}}, \bibinfo {author}
  {\bibfnamefont {Mark~T.}\ \bibnamefont {Edmonds}}, \bibinfo {author}
  {\bibfnamefont {Hsin-Zon}\ \bibnamefont {Tsai}}, \bibinfo {author}
  {\bibfnamefont {Alexander}\ \bibnamefont {Riss}}, \bibinfo {author}
  {\bibfnamefont {Sung-Kwan}\ \bibnamefont {Mo}}, \bibinfo {author}
  {\bibfnamefont {Dunghai}\ \bibnamefont {Lee}}, \bibinfo {author}
  {\bibfnamefont {Alex}\ \bibnamefont {Zettl}}, \bibinfo {author}
  {\bibfnamefont {Zahid}\ \bibnamefont {Hussain}}, \bibinfo {author}
  {\bibfnamefont {Zhi-Xun}\ \bibnamefont {Shen}}, \ and\ \bibinfo {author}
  {\bibfnamefont {Michael~F.}\ \bibnamefont {Crommie}},\ }\bibfield  {title}
  {\enquote {\bibinfo {title} {Characterization of collective ground states in
  single-layer {N}b{S}e$_2$},}\ }\href {\doibase 10.1038/nphys3527} {\bibfield
  {journal} {\bibinfo  {journal} {Nature Physics}\ }\textbf {\bibinfo {volume}
  {12}},\ \bibinfo {pages} {92--97} (\bibinfo {year} {2016})}\BibitemShut
  {NoStop}%
\bibitem [{\citenamefont {Saito}\ \emph {et~al.}(2016)\citenamefont {Saito},
  \citenamefont {Nakamura}, \citenamefont {Bahramy}, \citenamefont {Kohama},
  \citenamefont {Ye}, \citenamefont {Kasahara}, \citenamefont {Nakagawa},
  \citenamefont {Onga}, \citenamefont {Tokunaga}, \citenamefont {Nojima},
  \citenamefont {Yanase},\ and\ \citenamefont {Iwasa}}]{Saito:16}%
  \BibitemOpen
  \bibfield  {author} {\bibinfo {author} {\bibfnamefont {Yu}~\bibnamefont
  {Saito}}, \bibinfo {author} {\bibfnamefont {Yasuharu}\ \bibnamefont
  {Nakamura}}, \bibinfo {author} {\bibfnamefont {Mohammad~Saeed}\ \bibnamefont
  {Bahramy}}, \bibinfo {author} {\bibfnamefont {Yoshimitsu}\ \bibnamefont
  {Kohama}}, \bibinfo {author} {\bibfnamefont {Jianting}\ \bibnamefont {Ye}},
  \bibinfo {author} {\bibfnamefont {Yuichi}\ \bibnamefont {Kasahara}}, \bibinfo
  {author} {\bibfnamefont {Yuji}\ \bibnamefont {Nakagawa}}, \bibinfo {author}
  {\bibfnamefont {Masaru}\ \bibnamefont {Onga}}, \bibinfo {author}
  {\bibfnamefont {Masashi}\ \bibnamefont {Tokunaga}}, \bibinfo {author}
  {\bibfnamefont {Tsutomu}\ \bibnamefont {Nojima}}, \bibinfo {author}
  {\bibfnamefont {Youichi}\ \bibnamefont {Yanase}}, \ and\ \bibinfo {author}
  {\bibfnamefont {Yoshihiro}\ \bibnamefont {Iwasa}},\ }\bibfield  {title}
  {\enquote {\bibinfo {title} {Superconductivity protected by spin--valley
  locking in ion-gated {MoS}$_2$},}\ }\href {\doibase 10.1038/nphys3580}
  {\bibfield  {journal} {\bibinfo  {journal} {Nature Physics}\ }\textbf
  {\bibinfo {volume} {12}},\ \bibinfo {pages} {144--149} (\bibinfo {year}
  {2016})}\BibitemShut {NoStop}%
\bibitem [{\citenamefont {Xi}\ \emph {et~al.}(2016)\citenamefont {Xi},
  \citenamefont {Wang}, \citenamefont {Zhao}, \citenamefont {Park},
  \citenamefont {Law}, \citenamefont {Berger}, \citenamefont {Forr{\'o}},
  \citenamefont {Shan},\ and\ \citenamefont {Mak}}]{Xi:16}%
  \BibitemOpen
  \bibfield  {author} {\bibinfo {author} {\bibfnamefont {Xiaoxiang}\
  \bibnamefont {Xi}}, \bibinfo {author} {\bibfnamefont {Zefang}\ \bibnamefont
  {Wang}}, \bibinfo {author} {\bibfnamefont {Weiwei}\ \bibnamefont {Zhao}},
  \bibinfo {author} {\bibfnamefont {Ju-Hyun}\ \bibnamefont {Park}}, \bibinfo
  {author} {\bibfnamefont {Kam~Tuen}\ \bibnamefont {Law}}, \bibinfo {author}
  {\bibfnamefont {Helmuth}\ \bibnamefont {Berger}}, \bibinfo {author}
  {\bibfnamefont {L{\'a}szl{\'o}}\ \bibnamefont {Forr{\'o}}}, \bibinfo {author}
  {\bibfnamefont {Jie}\ \bibnamefont {Shan}}, \ and\ \bibinfo {author}
  {\bibfnamefont {Kin~Fai}\ \bibnamefont {Mak}},\ }\bibfield  {title} {\enquote
  {\bibinfo {title} {Ising pairing in superconducting {NbSe}$_2$ atomic
  layers},}\ }\href {\doibase 10.1038/nphys3538} {\bibfield  {journal}
  {\bibinfo  {journal} {Nature Physics}\ }\textbf {\bibinfo {volume} {12}},\
  \bibinfo {pages} {139--143} (\bibinfo {year} {2016})}\BibitemShut {NoStop}%
\bibitem [{\citenamefont {Qi}\ \emph {et~al.}(2016)\citenamefont {Qi},
  \citenamefont {Naumov}, \citenamefont {Ali}, \citenamefont {Rajamathi},
  \citenamefont {Schnelle}, \citenamefont {Barkalov}, \citenamefont {Hanfland},
  \citenamefont {Wu}, \citenamefont {Shekhar}, \citenamefont {Sun},
  \citenamefont {S{\"u}{\ss}}, \citenamefont {Schmidt}, \citenamefont
  {Schwarz}, \citenamefont {Pippel}, \citenamefont {Werner}, \citenamefont
  {Hillebrand}, \citenamefont {F{\"o}rster}, \citenamefont {Kampert},
  \citenamefont {Parkin}, \citenamefont {Cava}, \citenamefont {Felser},
  \citenamefont {Yan},\ and\ \citenamefont {Medvedev}}]{Qi:16}%
  \BibitemOpen
  \bibfield  {author} {\bibinfo {author} {\bibfnamefont {Yanpeng}\ \bibnamefont
  {Qi}}, \bibinfo {author} {\bibfnamefont {Pavel~G.}\ \bibnamefont {Naumov}},
  \bibinfo {author} {\bibfnamefont {Mazhar~N.}\ \bibnamefont {Ali}}, \bibinfo
  {author} {\bibfnamefont {Catherine~R.}\ \bibnamefont {Rajamathi}}, \bibinfo
  {author} {\bibfnamefont {Walter}\ \bibnamefont {Schnelle}}, \bibinfo {author}
  {\bibfnamefont {Oleg}\ \bibnamefont {Barkalov}}, \bibinfo {author}
  {\bibfnamefont {Michael}\ \bibnamefont {Hanfland}}, \bibinfo {author}
  {\bibfnamefont {Shu-Chun}\ \bibnamefont {Wu}}, \bibinfo {author}
  {\bibfnamefont {Chandra}\ \bibnamefont {Shekhar}}, \bibinfo {author}
  {\bibfnamefont {Yan}\ \bibnamefont {Sun}}, \bibinfo {author} {\bibfnamefont
  {Vicky}\ \bibnamefont {S{\"u}{\ss}}}, \bibinfo {author} {\bibfnamefont
  {Marcus}\ \bibnamefont {Schmidt}}, \bibinfo {author} {\bibfnamefont {Ulrich}\
  \bibnamefont {Schwarz}}, \bibinfo {author} {\bibfnamefont {Eckhard}\
  \bibnamefont {Pippel}}, \bibinfo {author} {\bibfnamefont {Peter}\
  \bibnamefont {Werner}}, \bibinfo {author} {\bibfnamefont {Reinald}\
  \bibnamefont {Hillebrand}}, \bibinfo {author} {\bibfnamefont {Tobias}\
  \bibnamefont {F{\"o}rster}}, \bibinfo {author} {\bibfnamefont {Erik}\
  \bibnamefont {Kampert}}, \bibinfo {author} {\bibfnamefont {Stuart}\
  \bibnamefont {Parkin}}, \bibinfo {author} {\bibfnamefont {R.~J.}\
  \bibnamefont {Cava}}, \bibinfo {author} {\bibfnamefont {Claudia}\
  \bibnamefont {Felser}}, \bibinfo {author} {\bibfnamefont {Binghai}\
  \bibnamefont {Yan}}, \ and\ \bibinfo {author} {\bibfnamefont {Sergey~A.}\
  \bibnamefont {Medvedev}},\ }\bibfield  {title} {\enquote {\bibinfo {title}
  {Superconductivity in {Weyl} semimetal candidate {M}o{T}e$_2$},}\ }\href
  {\doibase 10.1038/ncomms11038} {\bibfield  {journal} {\bibinfo  {journal}
  {Nature Communications}\ }\textbf {\bibinfo {volume} {7}},\ \bibinfo {pages}
  {11038} (\bibinfo {year} {2016})}\BibitemShut {NoStop}%
\bibitem [{\citenamefont {Costanzo}\ \emph {et~al.}(2016)\citenamefont
  {Costanzo}, \citenamefont {Jo}, \citenamefont {Berger},\ and\ \citenamefont
  {Morpurgo}}]{Costanzo:16}%
  \BibitemOpen
  \bibfield  {author} {\bibinfo {author} {\bibfnamefont {Davide}\ \bibnamefont
  {Costanzo}}, \bibinfo {author} {\bibfnamefont {Sanghyun}\ \bibnamefont {Jo}},
  \bibinfo {author} {\bibfnamefont {Helmuth}\ \bibnamefont {Berger}}, \ and\
  \bibinfo {author} {\bibfnamefont {Alberto~F.}\ \bibnamefont {Morpurgo}},\
  }\bibfield  {title} {\enquote {\bibinfo {title} {Gate-induced
  superconductivity in atomically thin {MoS}$_2$ crystals},}\ }\href {\doibase
  10.1038/nnano.2015.314} {\bibfield  {journal} {\bibinfo  {journal} {Nature
  Nanotechnology}\ }\textbf {\bibinfo {volume} {11}},\ \bibinfo {pages}
  {339--344} (\bibinfo {year} {2016})}\BibitemShut {NoStop}%
\bibitem [{\citenamefont {Dvir}\ \emph {et~al.}(2018)\citenamefont {Dvir},
  \citenamefont {Massee}, \citenamefont {Attias}, \citenamefont {Khodas},
  \citenamefont {Aprili}, \citenamefont {Quay},\ and\ \citenamefont
  {Steinberg}}]{Dvir:18}%
  \BibitemOpen
  \bibfield  {author} {\bibinfo {author} {\bibfnamefont {T.}~\bibnamefont
  {Dvir}}, \bibinfo {author} {\bibfnamefont {F.}~\bibnamefont {Massee}},
  \bibinfo {author} {\bibfnamefont {L.}~\bibnamefont {Attias}}, \bibinfo
  {author} {\bibfnamefont {M.}~\bibnamefont {Khodas}}, \bibinfo {author}
  {\bibfnamefont {M.}~\bibnamefont {Aprili}}, \bibinfo {author} {\bibfnamefont
  {C.~H.~L.}\ \bibnamefont {Quay}}, \ and\ \bibinfo {author} {\bibfnamefont
  {H.}~\bibnamefont {Steinberg}},\ }\bibfield  {title} {\enquote {\bibinfo
  {title} {Spectroscopy of bulk and few-layer superconducting {NbSe}$_2$ with
  van der waals tunnel junctions},}\ }\href {\doibase
  10.1038/s41467-018-03000-w} {\bibfield  {journal} {\bibinfo  {journal}
  {Nature Communications}\ }\textbf {\bibinfo {volume} {9}},\ \bibinfo {pages}
  {598} (\bibinfo {year} {2018})}\BibitemShut {NoStop}%
\bibitem [{\citenamefont {de~la Barrera}\ \emph {et~al.}(2018)\citenamefont
  {de~la Barrera}, \citenamefont {Sinko}, \citenamefont {Gopalan},
  \citenamefont {Sivadas}, \citenamefont {Seyler}, \citenamefont {Watanabe},
  \citenamefont {Taniguchi}, \citenamefont {Tsen}, \citenamefont {Xu},
  \citenamefont {Xiao},\ and\ \citenamefont {Hunt}}]{Barrera:18}%
  \BibitemOpen
  \bibfield  {author} {\bibinfo {author} {\bibfnamefont {Sergio~C.}\
  \bibnamefont {de~la Barrera}}, \bibinfo {author} {\bibfnamefont {Michael~R.}\
  \bibnamefont {Sinko}}, \bibinfo {author} {\bibfnamefont {Devashish~P.}\
  \bibnamefont {Gopalan}}, \bibinfo {author} {\bibfnamefont {Nikhil}\
  \bibnamefont {Sivadas}}, \bibinfo {author} {\bibfnamefont {Kyle~L.}\
  \bibnamefont {Seyler}}, \bibinfo {author} {\bibfnamefont {Kenji}\
  \bibnamefont {Watanabe}}, \bibinfo {author} {\bibfnamefont {Takashi}\
  \bibnamefont {Taniguchi}}, \bibinfo {author} {\bibfnamefont {Adam~W.}\
  \bibnamefont {Tsen}}, \bibinfo {author} {\bibfnamefont {Xiaodong}\
  \bibnamefont {Xu}}, \bibinfo {author} {\bibfnamefont {Di}~\bibnamefont
  {Xiao}}, \ and\ \bibinfo {author} {\bibfnamefont {Benjamin~M.}\ \bibnamefont
  {Hunt}},\ }\bibfield  {title} {\enquote {\bibinfo {title} {Tuning ising
  superconductivity with layer and spin--orbit coupling in two-dimensional
  transition-metal dichalcogenides},}\ }\href {\doibase
  10.1038/s41467-018-03888-4} {\bibfield  {journal} {\bibinfo  {journal}
  {Nature Communications}\ }\textbf {\bibinfo {volume} {9}},\ \bibinfo {pages}
  {1427} (\bibinfo {year} {2018})}\BibitemShut {NoStop}%
\bibitem [{\citenamefont {Sohn}\ \emph {et~al.}(2018)\citenamefont {Sohn},
  \citenamefont {Xi}, \citenamefont {He}, \citenamefont {Jiang}, \citenamefont
  {Wang}, \citenamefont {Kang}, \citenamefont {Park}, \citenamefont {Berger},
  \citenamefont {Forr{\'o}}, \citenamefont {Law}, \citenamefont {Shan},\ and\
  \citenamefont {Mak}}]{Sohn:18}%
  \BibitemOpen
  \bibfield  {author} {\bibinfo {author} {\bibfnamefont {Egon}\ \bibnamefont
  {Sohn}}, \bibinfo {author} {\bibfnamefont {Xiaoxiang}\ \bibnamefont {Xi}},
  \bibinfo {author} {\bibfnamefont {Wen-Yu}\ \bibnamefont {He}}, \bibinfo
  {author} {\bibfnamefont {Shengwei}\ \bibnamefont {Jiang}}, \bibinfo {author}
  {\bibfnamefont {Zefang}\ \bibnamefont {Wang}}, \bibinfo {author}
  {\bibfnamefont {Kaifei}\ \bibnamefont {Kang}}, \bibinfo {author}
  {\bibfnamefont {Ju-Hyun}\ \bibnamefont {Park}}, \bibinfo {author}
  {\bibfnamefont {Helmuth}\ \bibnamefont {Berger}}, \bibinfo {author}
  {\bibfnamefont {L{\'a}szl{\'o}}\ \bibnamefont {Forr{\'o}}}, \bibinfo {author}
  {\bibfnamefont {Kam~Tuen}\ \bibnamefont {Law}}, \bibinfo {author}
  {\bibfnamefont {Jie}\ \bibnamefont {Shan}}, \ and\ \bibinfo {author}
  {\bibfnamefont {Kin~Fai}\ \bibnamefont {Mak}},\ }\bibfield  {title} {\enquote
  {\bibinfo {title} {An unusual continuous paramagnetic-limited superconducting
  phase transition in {2D} {NbSe}$_2$},}\ }\href {\doibase
  10.1038/s41563-018-0061-1} {\bibfield  {journal} {\bibinfo  {journal} {Nature
  Materials}\ }\textbf {\bibinfo {volume} {17}},\ \bibinfo {pages} {504--508}
  (\bibinfo {year} {2018})}\BibitemShut {NoStop}%
\bibitem [{\citenamefont {Sajadi}\ \emph {et~al.}(2018)\citenamefont {Sajadi},
  \citenamefont {Palomaki}, \citenamefont {Fei}, \citenamefont {Zhao},
  \citenamefont {Bement}, \citenamefont {Olsen}, \citenamefont {Luescher},
  \citenamefont {Xu}, \citenamefont {Folk},\ and\ \citenamefont
  {Cobden}}]{Cobden:18}%
  \BibitemOpen
  \bibfield  {author} {\bibinfo {author} {\bibfnamefont {Ebrahim}\ \bibnamefont
  {Sajadi}}, \bibinfo {author} {\bibfnamefont {Tauno}\ \bibnamefont
  {Palomaki}}, \bibinfo {author} {\bibfnamefont {Zaiyao}\ \bibnamefont {Fei}},
  \bibinfo {author} {\bibfnamefont {Wenjin}\ \bibnamefont {Zhao}}, \bibinfo
  {author} {\bibfnamefont {Philip}\ \bibnamefont {Bement}}, \bibinfo {author}
  {\bibfnamefont {Christian}\ \bibnamefont {Olsen}}, \bibinfo {author}
  {\bibfnamefont {Silvia}\ \bibnamefont {Luescher}}, \bibinfo {author}
  {\bibfnamefont {Xiaodong}\ \bibnamefont {Xu}}, \bibinfo {author}
  {\bibfnamefont {Joshua~A.}\ \bibnamefont {Folk}}, \ and\ \bibinfo {author}
  {\bibfnamefont {David~H.}\ \bibnamefont {Cobden}},\ }\bibfield  {title}
  {\enquote {\bibinfo {title} {Gate-induced superconductivity in a monolayer
  topological insulator},}\ }\href {\doibase 10.1126/science.aar4426}
  {\bibfield  {journal} {\bibinfo  {journal} {Science}\ }\textbf {\bibinfo
  {volume} {362}},\ \bibinfo {pages} {922--925} (\bibinfo {year}
  {2018})}\BibitemShut {NoStop}%
\bibitem [{\citenamefont {Fatemi}\ \emph {et~al.}(2018)\citenamefont {Fatemi},
  \citenamefont {Wu}, \citenamefont {Cao}, \citenamefont {Bretheau},
  \citenamefont {Gibson}, \citenamefont {Watanabe}, \citenamefont {Taniguchi},
  \citenamefont {Cava},\ and\ \citenamefont {Jarillo-Herrero}}]{Fatemi:18}%
  \BibitemOpen
  \bibfield  {author} {\bibinfo {author} {\bibfnamefont {Valla}\ \bibnamefont
  {Fatemi}}, \bibinfo {author} {\bibfnamefont {Sanfeng}\ \bibnamefont {Wu}},
  \bibinfo {author} {\bibfnamefont {Yuan}\ \bibnamefont {Cao}}, \bibinfo
  {author} {\bibfnamefont {Landry}\ \bibnamefont {Bretheau}}, \bibinfo {author}
  {\bibfnamefont {Quinn~D.}\ \bibnamefont {Gibson}}, \bibinfo {author}
  {\bibfnamefont {Kenji}\ \bibnamefont {Watanabe}}, \bibinfo {author}
  {\bibfnamefont {Takashi}\ \bibnamefont {Taniguchi}}, \bibinfo {author}
  {\bibfnamefont {Robert~J.}\ \bibnamefont {Cava}}, \ and\ \bibinfo {author}
  {\bibfnamefont {Pablo}\ \bibnamefont {Jarillo-Herrero}},\ }\bibfield  {title}
  {\enquote {\bibinfo {title} {Electrically tunable low-density
  superconductivity in a monolayer topological insulator},}\ }\href {\doibase
  10.1126/science.aar4642} {\bibfield  {journal} {\bibinfo  {journal}
  {Science}\ }\textbf {\bibinfo {volume} {362}},\ \bibinfo {pages} {926--929}
  (\bibinfo {year} {2018})}\BibitemShut {NoStop}%
\bibitem [{\citenamefont {Hamill}\ \emph {et~al.}(2021)\citenamefont {Hamill},
  \citenamefont {Heischmidt}, \citenamefont {Sohn}, \citenamefont {Shaffer},
  \citenamefont {Tsai}, \citenamefont {Zhang}, \citenamefont {Xi},
  \citenamefont {Suslov}, \citenamefont {Berger}, \citenamefont {Forr{\'o}},
  \citenamefont {Burnell}, \citenamefont {Shan}, \citenamefont {Mak},
  \citenamefont {Fernandes}, \citenamefont {Wang},\ and\ \citenamefont
  {Pribiag}}]{Hamill:21}%
  \BibitemOpen
  \bibfield  {author} {\bibinfo {author} {\bibfnamefont {Alex}\ \bibnamefont
  {Hamill}}, \bibinfo {author} {\bibfnamefont {Brett}\ \bibnamefont
  {Heischmidt}}, \bibinfo {author} {\bibfnamefont {Egon}\ \bibnamefont {Sohn}},
  \bibinfo {author} {\bibfnamefont {Daniel}\ \bibnamefont {Shaffer}}, \bibinfo
  {author} {\bibfnamefont {Kan-Ting}\ \bibnamefont {Tsai}}, \bibinfo {author}
  {\bibfnamefont {Xi}~\bibnamefont {Zhang}}, \bibinfo {author} {\bibfnamefont
  {Xiaoxiang}\ \bibnamefont {Xi}}, \bibinfo {author} {\bibfnamefont {Alexey}\
  \bibnamefont {Suslov}}, \bibinfo {author} {\bibfnamefont {Helmuth}\
  \bibnamefont {Berger}}, \bibinfo {author} {\bibfnamefont {L{\'a}szl{\'o}}\
  \bibnamefont {Forr{\'o}}}, \bibinfo {author} {\bibfnamefont {Fiona~J.}\
  \bibnamefont {Burnell}}, \bibinfo {author} {\bibfnamefont {Jie}\ \bibnamefont
  {Shan}}, \bibinfo {author} {\bibfnamefont {Kin~Fai}\ \bibnamefont {Mak}},
  \bibinfo {author} {\bibfnamefont {Rafael~M.}\ \bibnamefont {Fernandes}},
  \bibinfo {author} {\bibfnamefont {Ke}~\bibnamefont {Wang}}, \ and\ \bibinfo
  {author} {\bibfnamefont {Vlad~S.}\ \bibnamefont {Pribiag}},\ }\bibfield
  {title} {\enquote {\bibinfo {title} {Two-fold symmetric superconductivity in
  few-layer {NbSe}$_2$},}\ }\href {\doibase 10.1038/s41567-021-01219-x}
  {\bibfield  {journal} {\bibinfo  {journal} {Nature Physics}\ }\textbf
  {\bibinfo {volume} {17}},\ \bibinfo {pages} {949--954} (\bibinfo {year}
  {2021})}\BibitemShut {NoStop}%
\bibitem [{\citenamefont {Xia}\ \emph {et~al.}(2025)\citenamefont {Xia},
  \citenamefont {Han}, \citenamefont {Watanabe}, \citenamefont {Taniguchi},
  \citenamefont {Shan},\ and\ \citenamefont {Mak}}]{Xia:25}%
  \BibitemOpen
  \bibfield  {author} {\bibinfo {author} {\bibfnamefont {Yiyu}\ \bibnamefont
  {Xia}}, \bibinfo {author} {\bibfnamefont {Zhongdong}\ \bibnamefont {Han}},
  \bibinfo {author} {\bibfnamefont {Kenji}\ \bibnamefont {Watanabe}}, \bibinfo
  {author} {\bibfnamefont {Takashi}\ \bibnamefont {Taniguchi}}, \bibinfo
  {author} {\bibfnamefont {Jie}\ \bibnamefont {Shan}}, \ and\ \bibinfo {author}
  {\bibfnamefont {Kin~Fai}\ \bibnamefont {Mak}},\ }\bibfield  {title} {\enquote
  {\bibinfo {title} {Superconductivity in twisted bilayer {WS}e$_2$},}\ }\href
  {\doibase 10.1038/s41586-024-08116-2} {\bibfield  {journal} {\bibinfo
  {journal} {Nature}\ }\textbf {\bibinfo {volume} {637}},\ \bibinfo {pages}
  {833--838} (\bibinfo {year} {2025})}\BibitemShut {NoStop}%
\bibitem [{\citenamefont {Nadeem}\ \emph {et~al.}(2023)\citenamefont {Nadeem},
  \citenamefont {Fuhrer},\ and\ \citenamefont {Wang}}]{Nadeem:23}%
  \BibitemOpen
  \bibfield  {author} {\bibinfo {author} {\bibfnamefont {Muhammad}\
  \bibnamefont {Nadeem}}, \bibinfo {author} {\bibfnamefont {Michael~S.}\
  \bibnamefont {Fuhrer}}, \ and\ \bibinfo {author} {\bibfnamefont {Xiaolin}\
  \bibnamefont {Wang}},\ }\bibfield  {title} {\enquote {\bibinfo {title} {The
  superconducting diode effect},}\ }\href {\doibase 10.1038/s42254-023-00632-w}
  {\bibfield  {journal} {\bibinfo  {journal} {Nature Reviews Physics}\ }\textbf
  {\bibinfo {volume} {5}},\ \bibinfo {pages} {558--577} (\bibinfo {year}
  {2023})}\BibitemShut {NoStop}%
\bibitem [{\citenamefont {Ma}\ \emph {et~al.}(2025)\citenamefont {Ma},
  \citenamefont {Zhan},\ and\ \citenamefont {Lin}}]{ma:25}%
  \BibitemOpen
  \bibfield  {author} {\bibinfo {author} {\bibfnamefont {Jiajun}\ \bibnamefont
  {Ma}}, \bibinfo {author} {\bibfnamefont {Ruiya}\ \bibnamefont {Zhan}}, \ and\
  \bibinfo {author} {\bibfnamefont {Xiao}\ \bibnamefont {Lin}},\ }\bibfield
  {title} {\enquote {\bibinfo {title} {Superconducting diode effects:
  Mechanisms, materials and applications},}\ }\href {\doibase
  https://doi.org/10.1002/apxr.202400180} {\bibfield  {journal} {\bibinfo
  {journal} {Advanced Physics Research}\ }\textbf {\bibinfo {volume} {4}},\
  \bibinfo {pages} {2400180} (\bibinfo {year} {2025})}\BibitemShut {NoStop}%
\bibitem [{\citenamefont {Shaffer}\ and\ \citenamefont
  {Levchenko}(2025)}]{Shaffer:25}%
  \BibitemOpen
  \bibfield  {author} {\bibinfo {author} {\bibfnamefont {Daniel}\ \bibnamefont
  {Shaffer}}\ and\ \bibinfo {author} {\bibfnamefont {Alex}\ \bibnamefont
  {Levchenko}},\ }\href {https://arxiv.org/abs/2510.25864} {\enquote {\bibinfo
  {title} {Theories of superconducting diode effects},}\ } (\bibinfo {year}
  {2025}),\ \Eprint {http://arxiv.org/abs/2510.25864} {arXiv:2510.25864
  [cond-mat.supr-con]} \BibitemShut {NoStop}%
\bibitem [{\citenamefont {Tokura}\ and\ \citenamefont
  {Nagaosa}(2018)}]{Tokura:18}%
  \BibitemOpen
  \bibfield  {author} {\bibinfo {author} {\bibfnamefont {Yoshinori}\
  \bibnamefont {Tokura}}\ and\ \bibinfo {author} {\bibfnamefont {Naoto}\
  \bibnamefont {Nagaosa}},\ }\bibfield  {title} {\enquote {\bibinfo {title}
  {Nonreciprocal responses from non-centrosymmetric quantum materials},}\
  }\href {\doibase 10.1038/s41467-018-05759-4} {\bibfield  {journal} {\bibinfo
  {journal} {Nature Communications}\ }\textbf {\bibinfo {volume} {9}},\
  \bibinfo {pages} {3740} (\bibinfo {year} {2018})}\BibitemShut {NoStop}%
\bibitem [{\citenamefont {Nagaosa}\ and\ \citenamefont
  {Yanase}(2024)}]{Nagaosa:24}%
  \BibitemOpen
  \bibfield  {author} {\bibinfo {author} {\bibfnamefont {Naoto}\ \bibnamefont
  {Nagaosa}}\ and\ \bibinfo {author} {\bibfnamefont {Youichi}\ \bibnamefont
  {Yanase}},\ }\bibfield  {title} {\enquote {\bibinfo {title} {Nonreciprocal
  transport and optical phenomena in quantum materials},}\ }\href {\doibase
  https://doi.org/10.1146/annurev-conmatphys-032822-033734} {\bibfield
  {journal} {\bibinfo  {journal} {Annual Review of Condensed Matter Physics}\
  }\textbf {\bibinfo {volume} {15}},\ \bibinfo {pages} {63--83} (\bibinfo
  {year} {2024})}\BibitemShut {NoStop}%
\bibitem [{\citenamefont {Wakatsuki}\ \emph {et~al.}(2017)\citenamefont
  {Wakatsuki}, \citenamefont {Saito}, \citenamefont {Hoshino}, \citenamefont
  {Itahashi}, \citenamefont {Ideue}, \citenamefont {Ezawa}, \citenamefont
  {Iwasa},\ and\ \citenamefont {Nagaosa}}]{Wakatsuki:17}%
  \BibitemOpen
  \bibfield  {author} {\bibinfo {author} {\bibfnamefont {Ryohei}\ \bibnamefont
  {Wakatsuki}}, \bibinfo {author} {\bibfnamefont {Yu}~\bibnamefont {Saito}},
  \bibinfo {author} {\bibfnamefont {Shintaro}\ \bibnamefont {Hoshino}},
  \bibinfo {author} {\bibfnamefont {Yuki~M.}\ \bibnamefont {Itahashi}},
  \bibinfo {author} {\bibfnamefont {Toshiya}\ \bibnamefont {Ideue}}, \bibinfo
  {author} {\bibfnamefont {Motohiko}\ \bibnamefont {Ezawa}}, \bibinfo {author}
  {\bibfnamefont {Yoshihiro}\ \bibnamefont {Iwasa}}, \ and\ \bibinfo {author}
  {\bibfnamefont {Naoto}\ \bibnamefont {Nagaosa}},\ }\bibfield  {title}
  {\enquote {\bibinfo {title} {Nonreciprocal charge transport in
  noncentrosymmetric superconductors},}\ }\href {\doibase
  10.1126/sciadv.1602390} {\bibfield  {journal} {\bibinfo  {journal} {Science
  Advances}\ }\textbf {\bibinfo {volume} {3}},\ \bibinfo {pages} {e1602390}
  (\bibinfo {year} {2017})}\BibitemShut {NoStop}%
\bibitem [{\citenamefont {Wakatsuki}\ and\ \citenamefont
  {Nagaosa}(2018)}]{Wakatsuki:18}%
  \BibitemOpen
  \bibfield  {author} {\bibinfo {author} {\bibfnamefont {Ryohei}\ \bibnamefont
  {Wakatsuki}}\ and\ \bibinfo {author} {\bibfnamefont {Naoto}\ \bibnamefont
  {Nagaosa}},\ }\bibfield  {title} {\enquote {\bibinfo {title} {Nonreciprocal
  current in noncentrosymmetric {Rashba} superconductors},}\ }\href {\doibase
  10.1103/PhysRevLett.121.026601} {\bibfield  {journal} {\bibinfo  {journal}
  {Phys. Rev. Lett.}\ }\textbf {\bibinfo {volume} {121}},\ \bibinfo {pages}
  {026601} (\bibinfo {year} {2018})}\BibitemShut {NoStop}%
\bibitem [{\citenamefont {Hoshino}\ \emph {et~al.}(2018)\citenamefont
  {Hoshino}, \citenamefont {Wakatsuki}, \citenamefont {Hamamoto},\ and\
  \citenamefont {Nagaosa}}]{Hoshino:18}%
  \BibitemOpen
  \bibfield  {author} {\bibinfo {author} {\bibfnamefont {Shintaro}\
  \bibnamefont {Hoshino}}, \bibinfo {author} {\bibfnamefont {Ryohei}\
  \bibnamefont {Wakatsuki}}, \bibinfo {author} {\bibfnamefont {Keita}\
  \bibnamefont {Hamamoto}}, \ and\ \bibinfo {author} {\bibfnamefont {Naoto}\
  \bibnamefont {Nagaosa}},\ }\bibfield  {title} {\enquote {\bibinfo {title}
  {Nonreciprocal charge transport in two-dimensional noncentrosymmetric
  superconductors},}\ }\href {\doibase 10.1103/PhysRevB.98.054510} {\bibfield
  {journal} {\bibinfo  {journal} {Phys. Rev. B}\ }\textbf {\bibinfo {volume}
  {98}},\ \bibinfo {pages} {054510} (\bibinfo {year} {2018})}\BibitemShut
  {NoStop}%
\bibitem [{\citenamefont {Daido}\ and\ \citenamefont
  {Yanase}(2024)}]{Daido:24}%
  \BibitemOpen
  \bibfield  {author} {\bibinfo {author} {\bibfnamefont {Akito}\ \bibnamefont
  {Daido}}\ and\ \bibinfo {author} {\bibfnamefont {Youichi}\ \bibnamefont
  {Yanase}},\ }\bibfield  {title} {\enquote {\bibinfo {title} {Rectification
  and nonlinear {Hall} effect by fluctuating finite-momentum {Cooper} pairs},}\
  }\href {\doibase 10.1103/PhysRevResearch.6.L022009} {\bibfield  {journal}
  {\bibinfo  {journal} {Phys. Rev. Res.}\ }\textbf {\bibinfo {volume} {6}},\
  \bibinfo {pages} {L022009} (\bibinfo {year} {2024})}\BibitemShut {NoStop}%
\bibitem [{\citenamefont {Dong}\ \emph {et~al.}(2025)\citenamefont {Dong},
  \citenamefont {Yang},\ and\ \citenamefont {Zhang}}]{Dong:25}%
  \BibitemOpen
  \bibfield  {author} {\bibinfo {author} {\bibfnamefont {Zi-Hao}\ \bibnamefont
  {Dong}}, \bibinfo {author} {\bibfnamefont {Hui}\ \bibnamefont {Yang}}, \ and\
  \bibinfo {author} {\bibfnamefont {Yi}~\bibnamefont {Zhang}},\ }\bibfield
  {title} {\enquote {\bibinfo {title} {Enhanced nonlinear {Hall} effect by
  {Cooper} pairs near the superconducting phase transition},}\ }\href {\doibase
  10.1103/PhysRevB.111.155120} {\bibfield  {journal} {\bibinfo  {journal}
  {Phys. Rev. B}\ }\textbf {\bibinfo {volume} {111}},\ \bibinfo {pages}
  {155120} (\bibinfo {year} {2025})}\BibitemShut {NoStop}%
\bibitem [{\citenamefont {de~Miranda}\ \emph
  {et~al.}(2026{\natexlab{a}})\citenamefont {de~Miranda}, \citenamefont
  {Khodas},\ and\ \citenamefont {Levchenko}}]{JTM:26a}%
  \BibitemOpen
  \bibfield  {author} {\bibinfo {author} {\bibfnamefont {Joaquim~Telles}\
  \bibnamefont {de~Miranda}}, \bibinfo {author} {\bibfnamefont {Maxim}\
  \bibnamefont {Khodas}}, \ and\ \bibinfo {author} {\bibfnamefont {Alex}\
  \bibnamefont {Levchenko}},\ }\href {https://arxiv.org/abs/2606.05302}
  {\enquote {\bibinfo {title} {Magnetochiral anisotropy in strained
  superconducting transition metal dichalcogenides},}\ } (\bibinfo {year}
  {2026}{\natexlab{a}}),\ \Eprint {http://arxiv.org/abs/2606.05302}
  {arXiv:2606.05302 [cond-mat.supr-con]} \BibitemShut {NoStop}%
\bibitem [{\citenamefont {de~Miranda}\ \emph
  {et~al.}(2026{\natexlab{b}})\citenamefont {de~Miranda}, \citenamefont
  {Khodas},\ and\ \citenamefont {Levchenko}}]{JTM:26b}%
  \BibitemOpen
  \bibfield  {author} {\bibinfo {author} {\bibfnamefont {Joaquim~Telles}\
  \bibnamefont {de~Miranda}}, \bibinfo {author} {\bibfnamefont {Maxim}\
  \bibnamefont {Khodas}}, \ and\ \bibinfo {author} {\bibfnamefont {Alex}\
  \bibnamefont {Levchenko}},\ }\href {https://arxiv.org/abs/2606.19421}
  {\enquote {\bibinfo {title} {Electrical magnetochiral anisotropy in {Rashba}
  superconductors},}\ } (\bibinfo {year} {2026}{\natexlab{b}}),\ \Eprint
  {http://arxiv.org/abs/2606.19421} {arXiv:2606.19421 [cond-mat.supr-con]}
  \BibitemShut {NoStop}%
\bibitem [{\citenamefont {Wangsness}\ and\ \citenamefont
  {Bloch}(1953)}]{Bloch:53}%
  \BibitemOpen
  \bibfield  {author} {\bibinfo {author} {\bibfnamefont {R.~K.}\ \bibnamefont
  {Wangsness}}\ and\ \bibinfo {author} {\bibfnamefont {F.}~\bibnamefont
  {Bloch}},\ }\bibfield  {title} {\enquote {\bibinfo {title} {The dynamical
  theory of nuclear induction},}\ }\href {\doibase 10.1103/PhysRev.89.728}
  {\bibfield  {journal} {\bibinfo  {journal} {Phys. Rev.}\ }\textbf {\bibinfo
  {volume} {89}},\ \bibinfo {pages} {728--739} (\bibinfo {year}
  {1953})}\BibitemShut {NoStop}%
\bibitem [{\citenamefont {Redfield}(1957)}]{Redfield:57}%
  \BibitemOpen
  \bibfield  {author} {\bibinfo {author} {\bibfnamefont {A.~G.}\ \bibnamefont
  {Redfield}},\ }\bibfield  {title} {\enquote {\bibinfo {title} {On the theory
  of relaxation processes},}\ }\href {\doibase 10.1147/rd.11.0019} {\bibfield
  {journal} {\bibinfo  {journal} {IBM J. Res. Dev.}\ }\textbf {\bibinfo
  {volume} {1}},\ \bibinfo {pages} {19--31} (\bibinfo {year}
  {1957})}\BibitemShut {NoStop}%
\bibitem [{\citenamefont {Korringa}(1950)}]{Korringa:50}%
  \BibitemOpen
  \bibfield  {author} {\bibinfo {author} {\bibfnamefont {J.}~\bibnamefont
  {Korringa}},\ }\bibfield  {title} {\enquote {\bibinfo {title} {Nuclear
  magnetic relaxation and resonnance line shift in metals},}\ }\href {\doibase
  10.1016/0031-8914(50)90105-4} {\bibfield  {journal} {\bibinfo  {journal}
  {Physica}\ }\textbf {\bibinfo {volume} {16}},\ \bibinfo {pages} {601--610}
  (\bibinfo {year} {1950})}\BibitemShut {NoStop}%
\bibitem [{\citenamefont {Hebel}\ and\ \citenamefont
  {Slichter}(1957)}]{HebelSlichter:57}%
  \BibitemOpen
  \bibfield  {author} {\bibinfo {author} {\bibfnamefont {L.~C.}\ \bibnamefont
  {Hebel}}\ and\ \bibinfo {author} {\bibfnamefont {C.~P.}\ \bibnamefont
  {Slichter}},\ }\bibfield  {title} {\enquote {\bibinfo {title} {Nuclear
  relaxation in superconducting aluminum},}\ }\href {\doibase
  10.1103/PhysRev.107.901} {\bibfield  {journal} {\bibinfo  {journal} {Physical
  Review}\ }\textbf {\bibinfo {volume} {107}},\ \bibinfo {pages} {901--902}
  (\bibinfo {year} {1957})}\BibitemShut {NoStop}%
\bibitem [{\citenamefont {Larkin}\ and\ \citenamefont
  {Varlamov}(2005)}]{LarkinVarlamov:05}%
  \BibitemOpen
  \bibfield  {author} {\bibinfo {author} {\bibfnamefont {A.~I.}\ \bibnamefont
  {Larkin}}\ and\ \bibinfo {author} {\bibfnamefont {A.}~\bibnamefont
  {Varlamov}},\ }\href@noop {} {\emph {\bibinfo {title} {Theory of Fluctuations
  in Superconductors}}}\ (\bibinfo  {publisher} {OUP Oxford},\ \bibinfo {year}
  {2005})\BibitemShut {NoStop}%
\bibitem [{\citenamefont {Abrahams}\ \emph {et~al.}(1970)\citenamefont
  {Abrahams}, \citenamefont {Redi},\ and\ \citenamefont {Woo}}]{Abrahams:70}%
  \BibitemOpen
  \bibfield  {author} {\bibinfo {author} {\bibfnamefont {Elihu}\ \bibnamefont
  {Abrahams}}, \bibinfo {author} {\bibfnamefont {Martha}\ \bibnamefont {Redi}},
  \ and\ \bibinfo {author} {\bibfnamefont {James W.~F.}\ \bibnamefont {Woo}},\
  }\bibfield  {title} {\enquote {\bibinfo {title} {Effect of fluctuations on
  electronic properties above the superconducting transition},}\ }\href
  {\doibase 10.1103/PhysRevB.1.208} {\bibfield  {journal} {\bibinfo  {journal}
  {Phys. Rev. B}\ }\textbf {\bibinfo {volume} {1}},\ \bibinfo {pages}
  {208--213} (\bibinfo {year} {1970})}\BibitemShut {NoStop}%
\bibitem [{\citenamefont {Aslamasov}\ and\ \citenamefont
  {Larkin}(1968)}]{AL:68}%
  \BibitemOpen
  \bibfield  {author} {\bibinfo {author} {\bibfnamefont {L.G.}\ \bibnamefont
  {Aslamasov}}\ and\ \bibinfo {author} {\bibfnamefont {A.I.}\ \bibnamefont
  {Larkin}},\ }\bibfield  {title} {\enquote {\bibinfo {title} {The influence of
  fluctuation pairing of electrons on the conductivity of normal metal},}\
  }\href {\doibase https://doi.org/10.1016/0375-9601(68)90623-3} {\bibfield
  {journal} {\bibinfo  {journal} {Physics Letters A}\ }\textbf {\bibinfo
  {volume} {26}},\ \bibinfo {pages} {238--239} (\bibinfo {year}
  {1968})}\BibitemShut {NoStop}%
\bibitem [{\citenamefont {Maki}(1968)}]{Maki:68}%
  \BibitemOpen
  \bibfield  {author} {\bibinfo {author} {\bibfnamefont {Kazumi}\ \bibnamefont
  {Maki}},\ }\bibfield  {title} {\enquote {\bibinfo {title} {The critical
  fluctuation of the order parameter in type-{II} superconductors},}\ }\href
  {\doibase 10.1143/PTP.39.897} {\bibfield  {journal} {\bibinfo  {journal}
  {Progress of Theoretical Physics}\ }\textbf {\bibinfo {volume} {39}},\
  \bibinfo {pages} {897--906} (\bibinfo {year} {1968})}\BibitemShut {NoStop}%
\bibitem [{\citenamefont {Thompson}(1970)}]{Thompson:70}%
  \BibitemOpen
  \bibfield  {author} {\bibinfo {author} {\bibfnamefont {Richard~S.}\
  \bibnamefont {Thompson}},\ }\bibfield  {title} {\enquote {\bibinfo {title}
  {Microwave, flux flow, and fluctuation resistance of dirty type-{II}
  superconductors},}\ }\href {\doibase 10.1103/PhysRevB.1.327} {\bibfield
  {journal} {\bibinfo  {journal} {Phys. Rev. B}\ }\textbf {\bibinfo {volume}
  {1}},\ \bibinfo {pages} {327--333} (\bibinfo {year} {1970})}\BibitemShut
  {NoStop}%
\bibitem [{\citenamefont {Maniv}\ and\ \citenamefont
  {Alexander}(1976)}]{Maniv:76}%
  \BibitemOpen
  \bibfield  {author} {\bibinfo {author} {\bibfnamefont {Toshioka}\
  \bibnamefont {Maniv}}\ and\ \bibinfo {author} {\bibfnamefont {Shlomo}\
  \bibnamefont {Alexander}},\ }\bibfield  {title} {\enquote {\bibinfo {title}
  {Fluctuation effect on the nuclear magnetic relaxation in clean
  superconductors},}\ }\href {\doibase 10.1088/0022-3719/9/9/013} {\bibfield
  {journal} {\bibinfo  {journal} {Journal of Physics C: Solid State Physics}\
  }\textbf {\bibinfo {volume} {9}},\ \bibinfo {pages} {1699--1710} (\bibinfo
  {year} {1976})}\BibitemShut {NoStop}%
\bibitem [{\citenamefont {Kuboki}\ and\ \citenamefont
  {Fukuyama}(1989)}]{Kuboki:89}%
  \BibitemOpen
  \bibfield  {author} {\bibinfo {author} {\bibfnamefont {Kazuhiro}\
  \bibnamefont {Kuboki}}\ and\ \bibinfo {author} {\bibfnamefont {Hidetoshi}\
  \bibnamefont {Fukuyama}},\ }\bibfield  {title} {\enquote {\bibinfo {title}
  {Effects of superconducting fluctuations on nmr relaxation rate},}\
  }\href@noop {} {\bibfield  {journal} {\bibinfo  {journal} {J. Phys. Soc.
  Jpn.}\ }\textbf {\bibinfo {volume} {58}},\ \bibinfo {pages} {376} (\bibinfo
  {year} {1989})}\BibitemShut {NoStop}%
\bibitem [{\citenamefont {Heym}(1992)}]{Heym:92}%
  \BibitemOpen
  \bibfield  {author} {\bibinfo {author} {\bibfnamefont {J\"urgen}\
  \bibnamefont {Heym}},\ }\bibfield  {title} {\enquote {\bibinfo {title}
  {Effects of superconducting fluctuations on the {NMR} relaxation rate
  $t^{-1}_1$ of two-dimensional systems above $t_c$},}\ }\href@noop {}
  {\bibfield  {journal} {\bibinfo  {journal} {J. Low Temp. Phys.}\ }\textbf
  {\bibinfo {volume} {89}},\ \bibinfo {pages} {869} (\bibinfo {year}
  {1992})}\BibitemShut {NoStop}%
\bibitem [{\citenamefont {Randeria}\ and\ \citenamefont
  {Varlamov}(1994)}]{Randeria:94}%
  \BibitemOpen
  \bibfield  {author} {\bibinfo {author} {\bibfnamefont {Mohit}\ \bibnamefont
  {Randeria}}\ and\ \bibinfo {author} {\bibfnamefont {Andrei~A.}\ \bibnamefont
  {Varlamov}},\ }\bibfield  {title} {\enquote {\bibinfo {title} {Effect of
  superconducting fluctuations on spin susceptibility and {NMR} relaxation
  rate},}\ }\href {\doibase 10.1103/PhysRevB.50.10401} {\bibfield  {journal}
  {\bibinfo  {journal} {Phys. Rev. B}\ }\textbf {\bibinfo {volume} {50}},\
  \bibinfo {pages} {10401(R)--10404(R)} (\bibinfo {year} {1994})}\BibitemShut
  {NoStop}%
\bibitem [{\citenamefont {Glatz}\ \emph {et~al.}(2015)\citenamefont {Glatz},
  \citenamefont {Galda},\ and\ \citenamefont {Varlamov}}]{Glatz:15}%
  \BibitemOpen
  \bibfield  {author} {\bibinfo {author} {\bibfnamefont {A.}~\bibnamefont
  {Glatz}}, \bibinfo {author} {\bibfnamefont {A.}~\bibnamefont {Galda}}, \ and\
  \bibinfo {author} {\bibfnamefont {A.~A.}\ \bibnamefont {Varlamov}},\
  }\bibfield  {title} {\enquote {\bibinfo {title} {Effect of fluctuations on
  the {NMR} relaxation beyond the {Abrikosov} vortex state},}\ }\href {\doibase
  10.1103/PhysRevB.92.054513} {\bibfield  {journal} {\bibinfo  {journal} {Phys.
  Rev. B}\ }\textbf {\bibinfo {volume} {92}},\ \bibinfo {pages} {054513}
  (\bibinfo {year} {2015})}\BibitemShut {NoStop}%
\bibitem [{\citenamefont {Baranov}\ \emph {et~al.}(1993)\citenamefont
  {Baranov}, \citenamefont {Kagan},\ and\ \citenamefont {Mar'enko}}]{BKM:93}%
  \BibitemOpen
  \bibfield  {author} {\bibinfo {author} {\bibfnamefont {M.~A.}\ \bibnamefont
  {Baranov}}, \bibinfo {author} {\bibfnamefont {M.~Yu.}\ \bibnamefont {Kagan}},
  \ and\ \bibinfo {author} {\bibfnamefont {M.~S.}\ \bibnamefont {Mar'enko}},\
  }\bibfield  {title} {\enquote {\bibinfo {title} {Singularity in the
  quasiparticle interaction function in a {2D Fermi} gas},}\ }\href@noop {}
  {\bibfield  {journal} {\bibinfo  {journal} {JETP Lett.}\ }\textbf {\bibinfo
  {volume} {58}},\ \bibinfo {pages} {709} (\bibinfo {year} {1993})}\BibitemShut
  {NoStop}%
\bibitem [{\citenamefont {Belitz}\ \emph {et~al.}(1997)\citenamefont {Belitz},
  \citenamefont {Kirkpatrick},\ and\ \citenamefont {Vojta}}]{Belitz:97}%
  \BibitemOpen
  \bibfield  {author} {\bibinfo {author} {\bibfnamefont {D.}~\bibnamefont
  {Belitz}}, \bibinfo {author} {\bibfnamefont {T.~R.}\ \bibnamefont
  {Kirkpatrick}}, \ and\ \bibinfo {author} {\bibfnamefont {Thomas}\
  \bibnamefont {Vojta}},\ }\bibfield  {title} {\enquote {\bibinfo {title}
  {Nonanalytic behavior of the spin susceptibility in clean {Fermi} systems},}\
  }\href {\doibase 10.1103/PhysRevB.55.9452} {\bibfield  {journal} {\bibinfo
  {journal} {Phys. Rev. B}\ }\textbf {\bibinfo {volume} {55}},\ \bibinfo
  {pages} {9452--9462} (\bibinfo {year} {1997})}\BibitemShut {NoStop}%
\bibitem [{\citenamefont {Chitov}\ and\ \citenamefont
  {Millis}(2001)}]{Chitov:01}%
  \BibitemOpen
  \bibfield  {author} {\bibinfo {author} {\bibfnamefont {Gennady~Y.}\
  \bibnamefont {Chitov}}\ and\ \bibinfo {author} {\bibfnamefont {Andrew~J.}\
  \bibnamefont {Millis}},\ }\bibfield  {title} {\enquote {\bibinfo {title}
  {Leading temperature corrections to {Fermi}-liquid theory in two
  dimensions},}\ }\href {\doibase 10.1103/PhysRevLett.86.5337} {\bibfield
  {journal} {\bibinfo  {journal} {Phys. Rev. Lett.}\ }\textbf {\bibinfo
  {volume} {86}},\ \bibinfo {pages} {5337--5340} (\bibinfo {year}
  {2001})}\BibitemShut {NoStop}%
\bibitem [{\citenamefont {Chubukov}\ and\ \citenamefont
  {Maslov}(2004)}]{Chubukov:04}%
  \BibitemOpen
  \bibfield  {author} {\bibinfo {author} {\bibfnamefont {Andrey~V.}\
  \bibnamefont {Chubukov}}\ and\ \bibinfo {author} {\bibfnamefont {Dmitrii~L.}\
  \bibnamefont {Maslov}},\ }\bibfield  {title} {\enquote {\bibinfo {title}
  {Singular corrections to the {Fermi}-liquid theory},}\ }\href {\doibase
  10.1103/PhysRevB.69.121102} {\bibfield  {journal} {\bibinfo  {journal} {Phys.
  Rev. B}\ }\textbf {\bibinfo {volume} {69}},\ \bibinfo {pages} {121102(R)}
  (\bibinfo {year} {2004})}\BibitemShut {NoStop}%
\bibitem [{\citenamefont {Betouras}\ \emph {et~al.}(2005)\citenamefont
  {Betouras}, \citenamefont {Efremov},\ and\ \citenamefont
  {Chubukov}}]{Betouras:05}%
  \BibitemOpen
  \bibfield  {author} {\bibinfo {author} {\bibfnamefont {Joseph}\ \bibnamefont
  {Betouras}}, \bibinfo {author} {\bibfnamefont {Dmitri}\ \bibnamefont
  {Efremov}}, \ and\ \bibinfo {author} {\bibfnamefont {Andrey}\ \bibnamefont
  {Chubukov}},\ }\bibfield  {title} {\enquote {\bibinfo {title} {Thermodynamics
  of a {Fermi} liquid in a magnetic field},}\ }\href {\doibase
  10.1103/PhysRevB.72.115112} {\bibfield  {journal} {\bibinfo  {journal} {Phys.
  Rev. B}\ }\textbf {\bibinfo {volume} {72}},\ \bibinfo {pages} {115112}
  (\bibinfo {year} {2005})}\BibitemShut {NoStop}%
\bibitem [{\citenamefont {Schwiete}\ and\ \citenamefont
  {Efetov}(2006)}]{Schwiete:06}%
  \BibitemOpen
  \bibfield  {author} {\bibinfo {author} {\bibfnamefont {G.}~\bibnamefont
  {Schwiete}}\ and\ \bibinfo {author} {\bibfnamefont {K.~B.}\ \bibnamefont
  {Efetov}},\ }\bibfield  {title} {\enquote {\bibinfo {title} {Temperature
  dependence of the spin susceptibility of a clean {Fermi} gas with
  repulsion},}\ }\href {\doibase 10.1103/PhysRevB.74.165108} {\bibfield
  {journal} {\bibinfo  {journal} {Phys. Rev. B}\ }\textbf {\bibinfo {volume}
  {74}},\ \bibinfo {pages} {165108} (\bibinfo {year} {2006})}\BibitemShut
  {NoStop}%
\bibitem [{\citenamefont {Shekhter}\ and\ \citenamefont
  {Finkel'stein}(2006)}]{Shekhter:06}%
  \BibitemOpen
  \bibfield  {author} {\bibinfo {author} {\bibfnamefont {A.}~\bibnamefont
  {Shekhter}}\ and\ \bibinfo {author} {\bibfnamefont {A.~M.}\ \bibnamefont
  {Finkel'stein}},\ }\bibfield  {title} {\enquote {\bibinfo {title}
  {Temperature dependence of spin susceptibility in two-dimensional {Fermi}
  liquid systems},}\ }\href {\doibase 10.1103/PhysRevB.74.205122} {\bibfield
  {journal} {\bibinfo  {journal} {Phys. Rev. B}\ }\textbf {\bibinfo {volume}
  {74}},\ \bibinfo {pages} {205122} (\bibinfo {year} {2006})}\BibitemShut
  {NoStop}%
\bibitem [{\citenamefont {Rashba}(1959)}]{Rashba:59}%
  \BibitemOpen
  \bibfield  {author} {\bibinfo {author} {\bibfnamefont {E.~I.}\ \bibnamefont
  {Rashba}},\ }\bibfield  {title} {\enquote {\bibinfo {title} {Symmetry of
  bands in wurzite-type crystals. 1. symmetry of bands disregarding spin-orbit
  interaction},}\ }\href@noop {} {\bibfield  {journal} {\bibinfo  {journal}
  {Sov. Phys. Solid. State}\ }\textbf {\bibinfo {volume} {1}},\ \bibinfo
  {pages} {368} (\bibinfo {year} {1959})}\BibitemShut {NoStop}%
\bibitem [{\citenamefont {Bychkov}\ and\ \citenamefont
  {Rashba}(1984)}]{Bychkov:84}%
  \BibitemOpen
  \bibfield  {author} {\bibinfo {author} {\bibfnamefont {Yu.~A.}\ \bibnamefont
  {Bychkov}}\ and\ \bibinfo {author} {\bibfnamefont {E.~I.}\ \bibnamefont
  {Rashba}},\ }\bibfield  {title} {\enquote {\bibinfo {title} {Properties of a
  2{D} electron gas with lifted spectral degeneracy},}\ }\href@noop {}
  {\bibfield  {journal} {\bibinfo  {journal} {Sov. Phys. JETP Lett.}\ }\textbf
  {\bibinfo {volume} {39}},\ \bibinfo {pages} {66} (\bibinfo {year}
  {1984})}\BibitemShut {NoStop}%
\bibitem [{\citenamefont {Levitov}\ \emph
  {et~al.}(1985{\natexlab{a}})\citenamefont {Levitov}, \citenamefont
  {Nazarov},\ and\ \citenamefont {Eliashberg}}]{LNE:85a}%
  \BibitemOpen
  \bibfield  {author} {\bibinfo {author} {\bibfnamefont {L.~S.}\ \bibnamefont
  {Levitov}}, \bibinfo {author} {\bibfnamefont {Yu.~V.}\ \bibnamefont
  {Nazarov}}, \ and\ \bibinfo {author} {\bibfnamefont {G.~M.}\ \bibnamefont
  {Eliashberg}},\ }\bibfield  {title} {\enquote {\bibinfo {title}
  {Magnetoelectric effects in conductors with mirror isomer symmetry},}\
  }\href@noop {} {\bibfield  {journal} {\bibinfo  {journal} {Sov. Phys. -
  JETP}\ }\textbf {\bibinfo {volume} {61}},\ \bibinfo {pages} {133} (\bibinfo
  {year} {1985}{\natexlab{a}})}\BibitemShut {NoStop}%
\bibitem [{\citenamefont {Levitov}\ \emph
  {et~al.}(1985{\natexlab{b}})\citenamefont {Levitov}, \citenamefont
  {Nazarov},\ and\ \citenamefont {Eliashberg}}]{LNE:85b}%
  \BibitemOpen
  \bibfield  {author} {\bibinfo {author} {\bibfnamefont {L.~S.}\ \bibnamefont
  {Levitov}}, \bibinfo {author} {\bibfnamefont {Yu.~V.}\ \bibnamefont
  {Nazarov}}, \ and\ \bibinfo {author} {\bibfnamefont {G.~M.}\ \bibnamefont
  {Eliashberg}},\ }\bibfield  {title} {\enquote {\bibinfo {title} {Electron
  spin susceptibility of superconductors},}\ }\href@noop {} {\bibfield
  {journal} {\bibinfo  {journal} {JETP Lett.}\ }\textbf {\bibinfo {volume}
  {41}},\ \bibinfo {pages} {228} (\bibinfo {year}
  {1985}{\natexlab{b}})}\BibitemShut {NoStop}%
\bibitem [{\citenamefont {Edelstein}(1995)}]{Edelstein:95}%
  \BibitemOpen
  \bibfield  {author} {\bibinfo {author} {\bibfnamefont {Victor~M.}\
  \bibnamefont {Edelstein}},\ }\bibfield  {title} {\enquote {\bibinfo {title}
  {Magnetoelectric effect in polar superconductors},}\ }\href {\doibase
  10.1103/PhysRevLett.75.2004} {\bibfield  {journal} {\bibinfo  {journal}
  {Phys. Rev. Lett.}\ }\textbf {\bibinfo {volume} {75}},\ \bibinfo {pages}
  {2004--2007} (\bibinfo {year} {1995})}\BibitemShut {NoStop}%
\bibitem [{\citenamefont {Edelstein}(2005)}]{Edelstein:05}%
  \BibitemOpen
  \bibfield  {author} {\bibinfo {author} {\bibfnamefont {Victor~M.}\
  \bibnamefont {Edelstein}},\ }\bibfield  {title} {\enquote {\bibinfo {title}
  {Magnetoelectric effect in dirty superconductors with broken mirror
  symmetry},}\ }\href {\doibase 10.1103/PhysRevB.72.172501} {\bibfield
  {journal} {\bibinfo  {journal} {Phys. Rev. B}\ }\textbf {\bibinfo {volume}
  {72}},\ \bibinfo {pages} {172501} (\bibinfo {year} {2005})}\BibitemShut
  {NoStop}%
\bibitem [{\citenamefont {Frigeri}\ \emph {et~al.}(2004)\citenamefont
  {Frigeri}, \citenamefont {Agterberg},\ and\ \citenamefont
  {Sigrist}}]{Frigeri:04}%
  \BibitemOpen
  \bibfield  {author} {\bibinfo {author} {\bibfnamefont {P.~A.}\ \bibnamefont
  {Frigeri}}, \bibinfo {author} {\bibfnamefont {D.~F.}\ \bibnamefont
  {Agterberg}}, \ and\ \bibinfo {author} {\bibfnamefont {M.}~\bibnamefont
  {Sigrist}},\ }\bibfield  {title} {\enquote {\bibinfo {title} {Spin
  susceptibility in superconductors without inversion symmetry},}\ }\href
  {\doibase 10.1088/1367-2630/6/1/115} {\bibfield  {journal} {\bibinfo
  {journal} {New Journal of Physics}\ }\textbf {\bibinfo {volume} {6}},\
  \bibinfo {pages} {115} (\bibinfo {year} {2004})}\BibitemShut {NoStop}%
\bibitem [{\citenamefont {Hayashi}\ \emph {et~al.}(2006)\citenamefont
  {Hayashi}, \citenamefont {Wakabayashi}, \citenamefont {Frigeri},\ and\
  \citenamefont {Sigrist}}]{Hayashi:06}%
  \BibitemOpen
  \bibfield  {author} {\bibinfo {author} {\bibfnamefont {N.}~\bibnamefont
  {Hayashi}}, \bibinfo {author} {\bibfnamefont {K.}~\bibnamefont
  {Wakabayashi}}, \bibinfo {author} {\bibfnamefont {P.~A.}\ \bibnamefont
  {Frigeri}}, \ and\ \bibinfo {author} {\bibfnamefont {M.}~\bibnamefont
  {Sigrist}},\ }\bibfield  {title} {\enquote {\bibinfo {title} {Nuclear
  magnetic relaxation rate in a noncentrosymmetric superconductor},}\ }\href
  {\doibase 10.1103/PhysRevB.73.092508} {\bibfield  {journal} {\bibinfo
  {journal} {Phys. Rev. B}\ }\textbf {\bibinfo {volume} {73}},\ \bibinfo
  {pages} {092508} (\bibinfo {year} {2006})}\BibitemShut {NoStop}%
\bibitem [{\citenamefont {Edelstein}(2021)}]{Edelstein:21}%
  \BibitemOpen
  \bibfield  {author} {\bibinfo {author} {\bibfnamefont {Victor~M.}\
  \bibnamefont {Edelstein}},\ }\bibfield  {title} {\enquote {\bibinfo {title}
  {{Ginzburg-Landau} theory for impure superconductors of polar symmetry},}\
  }\href {\doibase 10.1103/PhysRevB.103.094507} {\bibfield  {journal} {\bibinfo
   {journal} {Phys. Rev. B}\ }\textbf {\bibinfo {volume} {103}},\ \bibinfo
  {pages} {094507} (\bibinfo {year} {2021})}\BibitemShut {NoStop}%
\end{thebibliography}%

\end{document}